\documentclass[letterpaper,11pt]{article}
\usepackage[margin=1in]{geometry}
\usepackage{setspace,url,hyperref,natbib}
\usepackage{amsmath,amsthm,amssymb,amsfonts}
\usepackage{graphicx}
\usepackage{float}
\usepackage{graphicx,natbib,url,float,subcaption,booktabs,multirow}
\newcommand{\N}{\ensuremath{\mathbb{N}}}
\newcommand{\R}{\ensuremath{\mathbb{R}}}
\newcommand{\I}{\ensuremath{\mathbb{I}}}

\begin{document}
\title{Penalized Principal Component Analysis Using Smoothing}
\author{Rebecca M.\ Hurwitz and Georg Hahn}
\date{Harvard T.H.\ Chan School of Public Health, Boston, MA 02115, USA\\
\bigskip
Corresponding author: rebeccahurwitz@hsph.harvard.edu}
\maketitle

\doublespacing
\abstract{Principal components computed via PCA (principal component analysis) are traditionally used to reduce dimensionality in genomic data or to correct for population stratification. In this paper, we explore the penalized eigenvalue problem (PEP) which reformulates the computation of the first eigenvector as an optimization problem and adds an $L_1$  penalty constraint to enforce sparseness of the solution. The contribution of our article is threefold. First, we extend PEP by applying smoothing to the original LASSO-type $L_1$ penalty. This allows one to compute analytical gradients which enable faster and more efficient minimization of the objective function associated with the optimization problem. Second, we demonstrate how higher order eigenvectors can be calculated with PEP using established results from singular value decomposition (SVD). Third, we present four experimental studies to demonstrate the usefulness of the smoothed penalized eigenvectors. Using data from the 1000 Genomes Project dataset, we empirically demonstrate that our proposed smoothed PEP allows one to increase numerical stability and obtain meaningful eigenvectors. We also employ the penalized eigenvector approach in two additional real data applications (computation of a polygenic risk score and clustering), demonstrating that exchanging the penalized eigenvectors for their smoothed counterparts can increase prediction accuracy in polygenic risk scores and enhance discernibility of clusterings. Moreover, we compare our proposed smoothed PEP to seven state-of-the-art algorithms for sparse PCA and evaluate the accuracy of the obtained eigenvectors, their support recovery, and their runtime.}


\section{Introduction}
\label{sec:introduction}
Statistical genomic analyses often utilize eigenvectors to adjust and correct for population stratification, or differences in frequencies of alleles between subgroups in a population. Eigenvectors are routinely used in principal component analysis (PCA) to transform large genomic datasets into reduced, lower-dimensional representations while aiming to preserve a maximal amount of information.

The classic eigenvector problem can easily be generalized to the so-called \textit{generalized eigenvector problem} (GEP), which asks to find a pair $(\lambda,x)$ satisfying $x \neq 0$ such that $Ax=\lambda Bx$ for two given matrices $A,B \in \R^{n \times n}$ and $n \in \N$. As GEP generalizes the usual eigenvector concept, its applications likewise cover multivariate analysis, PCA, canonical correlation analysis (CCA), linear discriminant analysis (LDA), or invariant coordinate selection \citep{Tyler2009, Li2007}. Moreover, GEP appears in nonlinear dimension reduction \citep{Kokiopoulou2011} and in computer
vision and image processing \citep{Zhang2013}. Naturally, several variants of GEP have been considered in the literature, for instance the $L_0$-constrained GEP \citep{Sriperumbudur2011}, numerical approximations \citep{Song2015}, the computation of GEP via maximization of the Rayleigh coefficient \citep{Tan2018}, the computation of multiple eigen-pairs \citep{Song2015}, or GEP under the assumption of positive definiteness \citep{Han2016}.

As noted in \cite{Jung2019}, sparsity-inducing methods for computing eigenvectors and factor loadings have the potential to improve estimation accuracy in settings of high dimensions and low sample size, thus providing better interpretable eigenvectors through variable selection. For this reason, in \cite{Witten2011} the authors introduce an optimization problem called the penalized eigenvalue problem (PEP). In PEP, the computation of the first eigenvector is expressed as the optimization of a quadratic form, where the maximum value of the objective function is attained for the first eigenvector. Such a formulation is interesting as it allows one to add a LASSO-type penalty \citep{Tibshirani1996} with $L_1$ loss to the objective function, thereby modifying the problem \citep{Gaynanova2019, Guan2022}. The tradeoff incurred by controlling the sparseness at the expense of changing the problem is one of the topics of the present work.

Solving the optimization problem posed by PEP comes with numerical issues which are caused by the non-differentiability of the $L_1$ penalty. In this work, we are interested in understanding the behavior of PEP when substituting the $L_1$ penalty with a smooth surrogate. The smooth surrogate is derived with the help of smoothing \citep{Chen1995, Nesterov2005, Trendafilov2021}, and it is guaranteed to be uniformly close to the original $L_1$ penalty while being differentiable everywhere. Moreover, we apply a computational trick involving an SVD decomposition in order to extract higher order eigenvectors.

We validate our approach in four experimental studies. First, we apply it to a dataset of the 1000 Genomes Project \citep{1000genomes, 1000genomesURL}, a catalogue of human genetic variation created with the goal of discerning common variants frequent in at least $1$ percent of the populations studied. To assess the quality of eigenvectors computed for cluster analysis, we compute within-cluster and between-cluster sums of squares as well as silhouette scores. Through those experiments, we demonstrate that our smoothing approach allows one to obtain numerically more stable eigenvectors. Second, we compare the penalized eigenvectors and the smoothed penalized eigenvectors in a polygenic risk score application which aims to predict SARS-CoV-2 mortality. We demonstrate that using the smoothed penalized eigenvectors yields a polygenic risk score with increased accuracy, measured using the area under the curve (AUC) metric, than using the unsmoothed counterparts. Third, we apply the unsmoothed and smoothed PEP to analyze clusters in the Iris benchmark dataset \citep{iris}, demonstrating that the principal components computed with smoothed PEP lead to an increased discernibility of the clustering. Fourth, we compare smoothed PEP with seven state-of-the-art algorithms for sparse PCA \citep{Journee2008, dAspremont2007, Jung2019, Gaynanova2017, Song2015, Zhang2010, Shen2008} on simulated data whereby we generate matrices with known planted eigenvectors and sparsity. We evaluate all approaches with respect to the accuracy of the obtained eigenvectors, their support recovery, and their runtime.

This article is structured as follows. Section~\ref{sec:methods} introduces both the penalized eigenvalue problem as well as smoothing methodology before presenting our proposed smoothed version of PEP. It also highlights our approach for computing higher order eigenvectors, discusses an iterative solving algorithm, and proposes a thresholding approach to enforce sparsity. All experiments are included in Section~\ref{sec:results}, where we showcase our experiments on the data of the 1000 Genomes Project, the polygenic risk score and clustering applications, as well as the simulation study with other state-of-the-art algorithms for sparse PCA. The article concludes with a discussion in Section~\ref{sec:discussion}. Two additional simulations can be found in Appendix~\ref{sec:additional_simulations}. 

The proposed methodology has been implemented in the R-package \textit{SPEV}, available on CRAN \citep{SPEV}. Throughout the article, the $L_1$ norm is denoted with $\Vert \cdot \Vert_1$, the Euclidean norm is denoted with $\Vert \cdot \Vert_2$, and $\I$ denotes the indicator function.
\section{Methods}
\label{sec:methods}
This section starts with a review of the penalized eigenvalue problem (Section~\ref{sec:penalizedEV}), introduced in \cite{Gaynanova2017}. Section~\ref{sec:review} reviews the smoothing of piecewise affine functions, which serves as the basis of the proposed smoothing approach for the penalized eigenvalue problem proposed in this publication. In Section~\ref{sec:smoothing}, we identify the non-smooth component of the penalized eigenvalue problem and apply smoothing to it. We conclude with a series of remarks on the extraction of higher-order eigenvectors besides the first eigenvector (Section~\ref{sec:SVD}), some considerations on an iterative solving scheme (Section~\ref{sec:iterative}), and thresholding to enforce sparsity (Section~\ref{sec:thresholding}).

\subsection{The penalized eigenvalue problem}
\label{sec:penalizedEV}
In order to allow for the addition of a penalty to the eigenvalue or eigenvector problem, we first formulate the computation of the leading eigenvector as an optimization problem \citep{Golub1996}. Precisely, letting $Q \in \R^{p \times p}$ be a symmetric and positive semidefinite matrix for a fixed dimension $p \in \N$, and $C \in \R^{p \times p}$ be a symmetric and strictly positive definite matrix, it can be shown that the solution of the optimization problem
\begin{align}
    v = \arg\max_{v \in \R^p} v^\top Q v \qquad \text{subject to} \quad v^\top C v \leq 1,
    \label{eq:optimization}
\end{align}
is the leading eigenvector of the matrix $C^{-1} Q$. Following this observation, the \textit{penalized eigenvalue problem} of \cite{Gaynanova2017} is defined by adding an $L_1$ penalty to the objective function being maximized in eq.~\eqref{eq:optimization}, leading to the Lasso-type \citep{Tibshirani1996} objective function
\begin{align}
    v_\lambda = \arg\max_{v:\Vert v \Vert=1} \left[ v^\top Q v - \lambda \Vert v \Vert_1 \right],
    \label{eq:pep}
\end{align}
where $C$ is assumed to be the identity matrix, and $\lambda>0$ is a penalty parameter. Note that in eq.~\eqref{eq:pep}, the objective is maximized, thus the penalty (which is non-negative) is being subtracted.

\subsection{A brief overview of the smoothing methodology}
\label{sec:review}
We employ the unified framework of \cite{Chen1995, Nesterov2005, Trendafilov2021} to smooth the objective function in eq.~\eqref{eq:pep}. We consider the smoothing of a piecewise affine and convex function $f:\R^q \rightarrow \R$, where $q \in \N$. As noted in \cite{Chen1995, Nesterov2005, Trendafilov2021}, piecewise affine functions can be written as
\begin{align}
    f(z) = \max_{i=1,\ldots,k} \left( A[z,1]^\top \right)_i,
    \label{eq:pwa}
\end{align}
where $z \in \R^q$ and $[z,1] \in \R^{q+1}$ denotes the concatenation of $z$ and the scalar $1$. The function in eq.~\eqref{eq:pwa} is called piecewise affine as it is composed of $k \in \N$ linear pieces, parameterized by the rows of $A \in \R^{k \times (q+1)}$. To be precise, each row in $A$ contains the coefficients of one of the $k$ linear pieces, with the constant coefficients being in column $q+1$.

A smooth surrogate for the function $f$ of eq.~\eqref{eq:pwa} is given by
\begin{align}
    f^\mu(z) = \max_{w \in Q_k} \left\{ \langle A[z,1]^\top,w \rangle - \mu \rho(w) \right\}.
    \label{eq:smoothed_pwa}
\end{align}
In eq.~\eqref{eq:smoothed_pwa}, the optimization is carried out over the unit simplex in $k$ dimensions, given by
$$Q_k = \left\{ w \in \R^k: \sum_{i=1}^k w_i=1, w_i \geq 0 ~\forall i=1,\ldots,k \right\} \subseteq \R^k.$$
The function $\rho$ which appears in eq.~\eqref{eq:smoothed_pwa} is called the prox-function. The prox-function must be nonnegative, continuously differentiable, and strongly convex. A specific choice of the prox-function is given below. The parameter $\mu>0$ is called the smoothing parameter. The choice $\mu=0$ recovers $f^0=f$.

The advantage of the smoothed surrogate of eq.~\eqref{eq:smoothed_pwa} consists of the fact that it is both smooth for any choice $\mu>0$ and uniformly close to $f$, that is the approximation error is uniformly bounded as
\begin{align}
    \sup_{z \in \R^q} \left| f(z)-f^\mu(z) \right| \leq \mu \sup_{w \in Q_k} \rho(w) = O(\mu),
    \label{eq:bound}
\end{align}
see \cite[Theorem~1]{Nesterov2005}. As can be seen from eq.~\eqref{eq:smoothed_pwa} and eq.~\eqref{eq:bound}, larger values of $\mu$ result in a stronger smoothing effect and a higher approximation error, while smaller values of $\mu$ result in a weaker smoothing effect and a higher degree of similarity between $f^\mu$ and $f$.

The choice of the prox-function $\rho$ is not unique, and in \cite{Chen1995, Nesterov2005, Trendafilov2021}, several choices for the prox-function can be found. In the remainder of the article, we focus on one particular choice called the entropy prox-function as it allows for simple closed-form expressions of the smoothed objective for the penalized eigenvalue problem of eq.~\eqref{eq:pep}. The entropy prox-function is given by
\begin{align}
    f_e^\mu(z) = \mu \log \left( \frac{1}{k} \sum_{i=1}^k e^\frac{ \left( A[z,1]^\top \right)_i }{\mu} \right),
    \label{eq:entropy_closedform}
\end{align}
and as shown in \cite[Section~4.1]{Nesterov2005} and \cite{smoothedLassoSTCO, framework}, it satisfies the uniform bound
\begin{align}
    \sup_{z \in \R^q} \left| f(z)-f_e^\mu(z) \right| \leq \mu \log(k)
    \label{eq:entropy_bound}
\end{align}
on the approximation error between $f^\mu$ and $f$.

\subsection{A smoothed version of the penalized eigenvalue problem}
\label{sec:smoothing}
Equipped with the results from Section~\ref{sec:review}, we now state the proposed smoothed version of the penalized eigenvalue problem of eq.~\eqref{eq:pep}. To this end, it is important to note that in eq.~\eqref{eq:pep}, the quadratic form $v^\top Q v$ is differentiable, while the $L_1$ penalty is not. We thus apply the smoothing to the $L_1$ penalty of the objective function only. Moreover, since $\Vert v \Vert_1$ in eq.~\eqref{eq:pep} for $v \in \R^p$ can be written as $\Vert v \Vert_1 = \sum_{i=1}^p |v_i|$, it suffices to smooth the non-differentiable absolute value applied to each entry of the vector $v$ separately.

The absolute value can be expressed in the form of eq.~\eqref{eq:pwa} with $k=2$ pieces, that is the function $f(z) = \max\{-z,z\} = \max_{i=1,2} \left( A[z,1]^\top \right)_i$ with
$$A = \left( \begin{array}{cc} -1 & 0\\1 & 0 \end{array}\right),$$
where $z \in \R$ is a scalar. We aim to replace the absolute value by the smooth surrogate with the entropy prox-function of eq.~\eqref{eq:entropy_closedform}. Simplifying eq.~\eqref{eq:entropy_closedform} for the specific choice of $A$ results in
\begin{align}
    f_e^\mu(z) = \mu \log \left( \frac{1}{2} e^{-z/\mu} + \frac{1}{2} e^{z/\mu} \right),
    \label{eq:smoothed_abs}
\end{align}
which is the smooth surrogate for the absolute value $|\cdot|$ we use. Leaving the quadratic form $v^\top Q v$ in eq.~\eqref{eq:pep} unchanged, and substituting the $L_1$ penalty $\Vert v \Vert_1 = \sum_{i=1}^p |v_i|$ by its smoothed version $\sum_{i=1}^p f_e^\mu(v_i)$, then results in the smoothed penalized eigenvalue problem
\begin{align}
    v_\lambda^\mu = \arg\max_{v:\Vert v \Vert=1} \left[ v^\top Q v - \lambda \sum_{i=1}^p f_e^\mu(v_i) \right].
    \label{eq:smoothed_pep}
\end{align}
Due to its simple form, the first derivative of $f_e^\mu$ can be stated explicitly as
$$\frac{\partial}{\partial z} f_e^\mu(z) = \frac{ -e^{-z/\mu}+e^{z/\mu} }{ e^{-z/\mu} + e^{z/\mu} } =: g_e^\mu(z).$$
Therefore, denoting the objective function in eq.~\eqref{eq:smoothed_pep} as
$$F_\lambda(v) = v^\top Q v - \lambda \sum_{i=1}^p f_e^\mu(v_i),$$
its gradient is given by
$$\nabla F_\lambda = 2Qv - \lambda \left[ g_e^\mu(v_1),\ldots,g_e^\mu(v_p) \right],$$
which is useful for the maximization of eq.~\eqref{eq:smoothed_pep} via gradient descent/ ascent solvers.

\subsection{Extraction of further eigenvectors via SVD}
\label{sec:SVD}
Solving the unsmoothed penalized eigenvalue problem of eq.~\eqref{eq:pep}, or the smoothed version of eq.~\eqref{eq:smoothed_pep}, yields the leading eigenvector. Naturally, we are interested in further eigenvectors, in particular in Section~\ref{sec:results} we will present population stratification plots of the first two leading eigenvectors, and compute a polygenic risk score using the first $10$ leading eigenvectors.

Deflation is a standard technique to extract higher order eigenvectors from a matrix $X \in \R^{m \times n}$, see \cite{Mackey2009, Trendafilov2021}. The idea is based on the Eckart-Young theorem \citep{EckartYoung1936} for the SVD representation $X = U \Sigma V$, where $U$ is an $m \times m$ unitary matrix, $\Sigma$ is an $m \times n$ matrix with off-diagonal elements (that is, $\Sigma_{ij}$ for $i \neq j$) being zero, and $V \in \R^{n \times n}$ is a unitary matrix. The representation can be rewritten as $X = \sum_i \sigma_i u_i \otimes v_i$, where $\otimes$ denotes the outer product between two vectors and $\sigma_i$ is the $i$th entry on the diagonal of $\Sigma$. Using this representation allows one to \textit{subtract} the leading eigenvector from $X$ once it is known.

We therefore first solve the unsmoothed penalized eigenvalue problem of eq.~\eqref{eq:pep}, or the smoothed version of eq.~\eqref{eq:smoothed_pep}, and obtain the leading eigenvector of a matrix $X$, denoted as $v_1$. We compute the corresponding eigenvalue using the Rayleigh coefficient as $\alpha_1 = v_1^\top X v_1$. We then compute $X_1 = X - \alpha_1 v_1 \otimes v_1$, the matrix with leading eigencomponent subtracted. As seen from the Eckart–Young theorem, the representation $X - \alpha_1 v_1 \otimes v_1 = \sum_{i>1} \alpha_i v_i \otimes v_i$ still holds true, now with the second eigencomponent being the leading one. Applying the (unsmoothed or smoothed) penalized eigenvalue problem to $X - \alpha_1 v_1 \otimes v_1$ thus yields the second penalized eigenvector.

\subsection{Iterative solving approach}
\label{sec:iterative}
Using the smoothed version of the penalized eigenvalue problem of eq.~\eqref{eq:smoothed_pep} and its explicit gradient $\nabla F_\lambda$ given in Section~\ref{sec:smoothing}, the minimization of eq.~\eqref{eq:smoothed_pep} is carried out using the quasi-Newton method \textit{BFGS} (Broyden--Fletcher--Goldfarb--Shanno) which is implemented in the function \textit{optim} in R \citep{RDevelopmentCoreTeam}.

To improve the numerical accuracy of optimizing eq.~\eqref{eq:smoothed_pep}, an iterative approach can be used. Given a target value of $\mu>0$ and an additional parameter $s \in \N$ (the number of steps), we start solving eq.~\eqref{eq:smoothed_pep} using a random initial starting value and the smoothing parameter $2^s\mu$. After the optimization is complete, we use the obtained maximizer as the initial value for a new run, this time with smoothing parameter $2^{s-1}\mu$. We continue seeding each new optimization with the solution of the previous run while lowering the smoothing parameter further until the target smoothing parameter $2^0\mu$ is reached. The target value of $\mu$ should be chosen as the machine precision or the square root of the machine precision. This is to ensure that the obtained estimates with smoothed PEP of eq.~\eqref{eq:smoothed_pep} closely resemble the ones obtained if the original PEP of eq.~\eqref{eq:pep} had been solved.

\subsection{Thresholding to enforce sparsity}
\label{sec:thresholding}
Solving the smoothed PEP objective of eq.~\eqref{eq:smoothed_pep} yields an eigenvector which is not guaranteed to be sparse. This is due to the lack of an $L_1$ penalty in the formulation of eq.~\eqref{eq:smoothed_pep}. Nevertheless, we can enforce sparsity again via thresholding. To be precise, let $v = v_\lambda^\mu$ be the vector obtained by maximizing eq.~\eqref{eq:smoothed_pep}. Denoting the vector as $v = (v_1,\dots,v_p)$, we first choose a desired sparsity level $\pi \in [0,1]$. We then compute the threshold $\tau$ as the empirical $\pi$-quantile of the values $\{ |v_1|, \ldots, |v_p| \}$, thus ensuring that a proportion $\pi$ of $v$ is smaller than $\tau$ in absolute value. Finally, we threshold all entries in $v$ against $\tau$, meaning we compute a new thresholded vector $\overline{v}$ with entries $\overline{v}_i = v_i \I(|v_i|>\tau)$ for all $i \in \{1,\ldots,p\}$.
\section{Experimental results}
\label{sec:results}
In this section, we present our numerical results, which are subdivided into several sections. Section~\ref{sec:setting} details the experimental setting and Section~\ref{sec:datasets} introduces the three datasets we employ, precisely the one of the 1000 Genomes Project \citep{1000genomes, 1000genomesURL}, the GISAID dataset \citep{Elbe2017, Shu2017} of SARS-CoV-2 nucleotide sequences, and the Iris dataset of the UC Irvine Machine Learning Repository \citep{iris}. The results for the unsmoothed penalized eigenvalue problem of eq.~\eqref{eq:pep} are presented in Section~\ref{sec:unsmooth} for a variety of Lasso penalties. The results for smoothed PEP are shown in Section~\ref{sec:smooth} using the same Lasso penalties. Moreover, we are interested in the quality of the clusters spanned by the eigenvectors in the PCA population stratification plots, and present cluster assessment metrics in Section~\ref{sec:clusters}. The choice of the smoothing parameter $\mu$ is investigated in Section~\ref{sec:mu}. Section~\ref{sec:prs} presents the first real data application, precisely the computation of a polygenic risk score to predict SARS-CoV-2 mortality using both unsmoothed and smoothed PEP eigenvectors. Section~\ref{sec:iris} presents the second real data application in the context of clustering in machine learning. Section~\ref{sec:comparison} presents the third simulation study in which we compare our smoothed PEP to seven other state-of-the-art algorithms.

\subsection{Experimental setting}
\label{sec:setting}
In all of our experiments, whenever we refer to the smoothed version of PEP, we use the iterative solving approach outlined in Section~\ref{sec:iterative} to solve eq.~\eqref{eq:smoothed_pep} with a target smoothing parameter of $\mu=0.1$. The smoothed version of PEP is used with $s=5$ steps, that is starting from $2^5\mu$. The choice of the parameter $\mu$ is investigated further in Section~\ref{sec:mu}.

Moreover, in all of our experiments, sparsity is enforced via thresholding as described in Section~\ref{sec:thresholding}. We use a target sparsity of $0.05$ (5\% percent zeros).

All experiments are carried out using the statistics software $R$ \citep{RDevelopmentCoreTeam}. The maximization of unsmoothed PEP in eq.~\eqref{eq:pep} is conducted with the FISTA algorithm of \cite{Beck2009}. The smoothed PEP of eq.~\eqref{eq:smoothed_pep} is carried out using the \textit{optim} function in $R$. The parameter $\lambda$ of eq.~\eqref{eq:pep} and eq.~\eqref{eq:smoothed_pep} is given individually in each simulation subsection.

In the comparison study of Section~\ref{sec:comparison}, we evaluate our proposed smoothed PEP to seven state-of-the-art sparse PCA algorithms. The details of those algorithms, including references to code we used, is given in Section~\ref{sec:comparison}.

\subsection{Datasets}
\label{sec:datasets}
The first dataset under consideration is the 1000 Genomes Project \citep{1000genomes, 1000genomesURL}. The goal of the 1000 Genomes Project was to discover genetic variants with frequencies of at least $1$ percent in the populations studied. In order to prepare the raw 1000 Genomes Project data and select rare variants, we used PLINK2 \citep{PurcellChang2019} with the cutoff value set to 0.01 for the {-}{-}\textit{max-maf} option. We then used the LD pruning method with the parameteters {-}{-}\textit{indep-pairwise} set to $2000$ $10$ $0.01$. These results focus on the European super population (subpopulations: CEU, FIN, GBR, IBS, and TSI), with our dataset containing 503 subjects and 5 million rare variants in total.

Next, we compute a similarity measure on the $503$ genomes included in the dataset. We employ the genomic relationship matrix (GRM) of \cite{Yang2011}. We then compute two sets of leading eigenvectors, precisely the first two eigenvectors of the unsmoothed (see eq.~\eqref{eq:pep}) and the smoothed (see eq.~\eqref{eq:smoothed_pep}) penalized eigenvector problems, respectively.

The second dataset under consideration consists of 1000 nucleotide sequences of the SARS-CoV-2 virus, which were extracted from patients diagnosed with Covid-19. The sequences were downloaded from the GISAID database \citep{Elbe2017, Shu2017}. The sequences have accession numbers in the range of EPI\_ISL\_404227 to EPI\_ISL\_766048. All nucleotide sequences were aligned with MAFFT \citep{Katoh2002} to the official SARS-CoV-2 reference sequence (available on GISAID under the accession number EPI\_ISL\_402124) using the \textit{keeplength} option, and with all other parameters set to their default values. This procedure yields aligned sequences of length $29891$ base pairs.

After alignment, we compare each aligned nucleotide sequence against the reference sequence, resulting in a binary vector with $1$ indicating a mismatch to the reference genome, and $0$ otherwise. As before, we compute the genomic relationship matrix (GRM) of \cite{Yang2011} on the binary matrix obtained in this way (with each row corresponding to one sample), resulting in a $1000 \times 1000$ similarity matrix for all subjects. The eigenvectors of the unsmoothed and smoothed PEP are computed on this similarity matrix (with $\lambda=0.1$ for smoothed PEP). Additional to the SARS-CoV-2 nucleotide sequence for each patient, we also obtain patient status meta information from GISAID, consisting of age (numeric), sex (male, female), geographic region (WHO geographic region encoded as \textit{WPRO, EURO, AFRO, PAHO, EMRO, SEARO}), and binary mortality (patient deceased which is encoded with 1, or alive which is encoded with 0).

The third dataset under consideration is the Iris dataset of UC Irvine Machine Learning Repository \citep{iris}, a simple but widely used benchmark for clustering. The dataset contains $150$ observations (rows), where each observation is a plant. The true labels is the plant species (setosa, versicolor, virginica). The covariates are the plant's sepal length, sepal width, petal length, and petal width (all measured as floating point numbers). The task is to recover the species from the four covariates measured for each plant.

\subsection{Results for the unsmoothed penalized eigenvectors}
\label{sec:unsmooth}
\begin{figure*}
    \centering
    \begin{subfigure}[b]{0.49\textwidth}
        \centering
        \includegraphics[width=\textwidth]{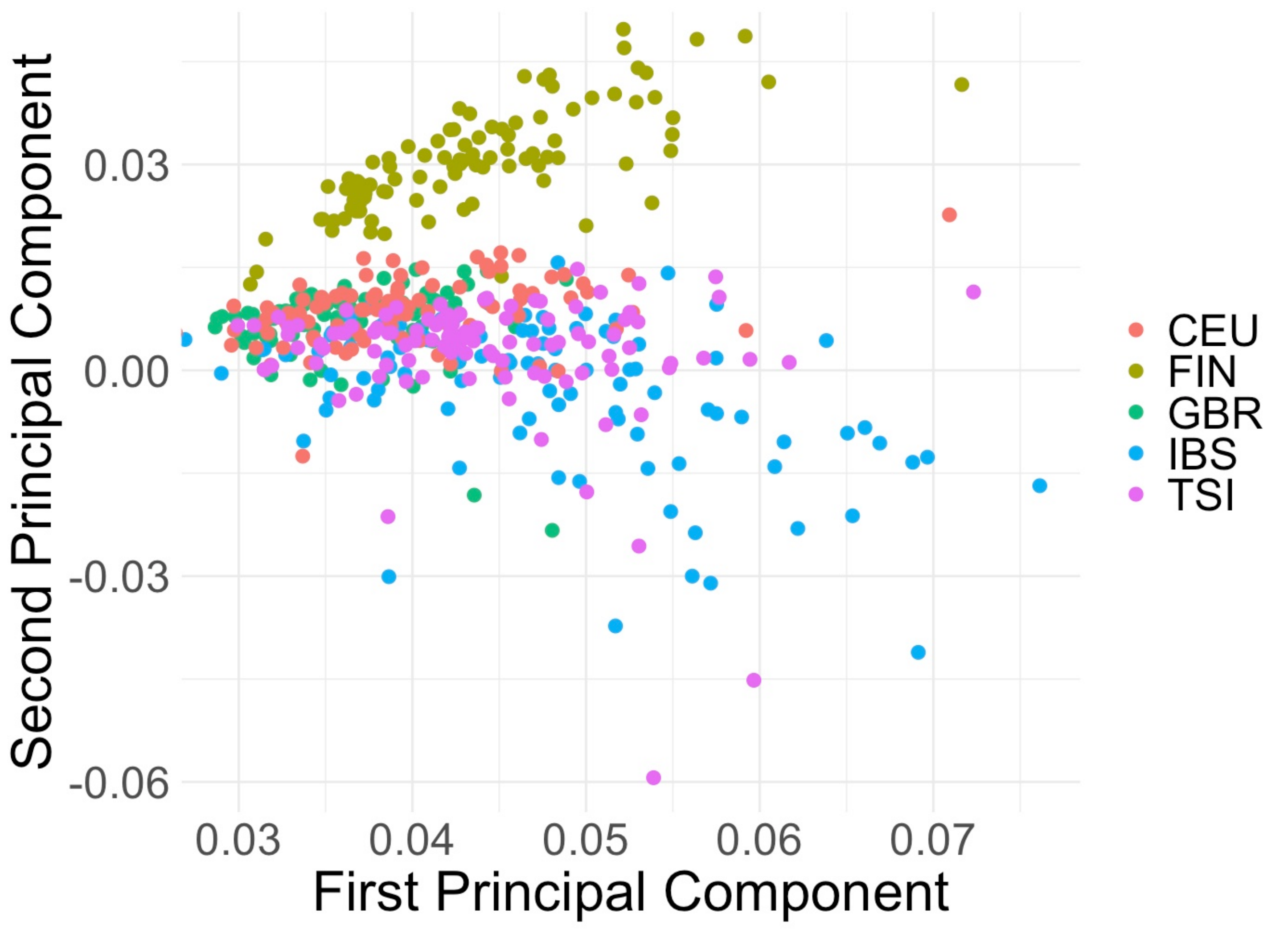}
        \caption{$\lambda=0$}
        \label{fig:unsmooth_a}
    \end{subfigure}
    \hfill
    \begin{subfigure}[b]{0.49\textwidth}
        \centering
        \includegraphics[width=\textwidth]{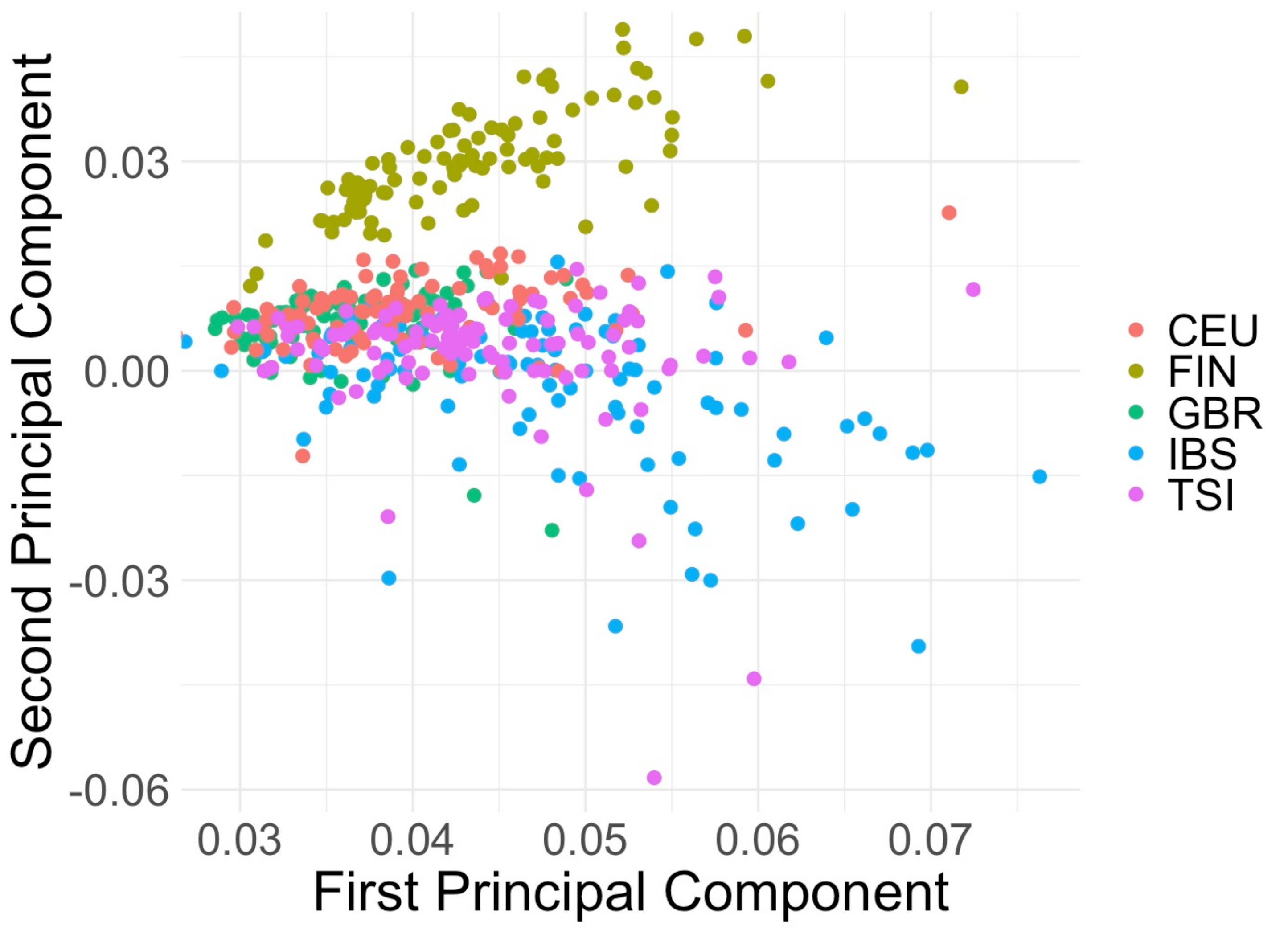}
        \caption{$\lambda=1$}
        \label{fig:unsmooth_b}
    \end{subfigure}
    
    \bigskip
    \begin{subfigure}[b]{0.49\textwidth}
        \centering
        \includegraphics[width=\textwidth]{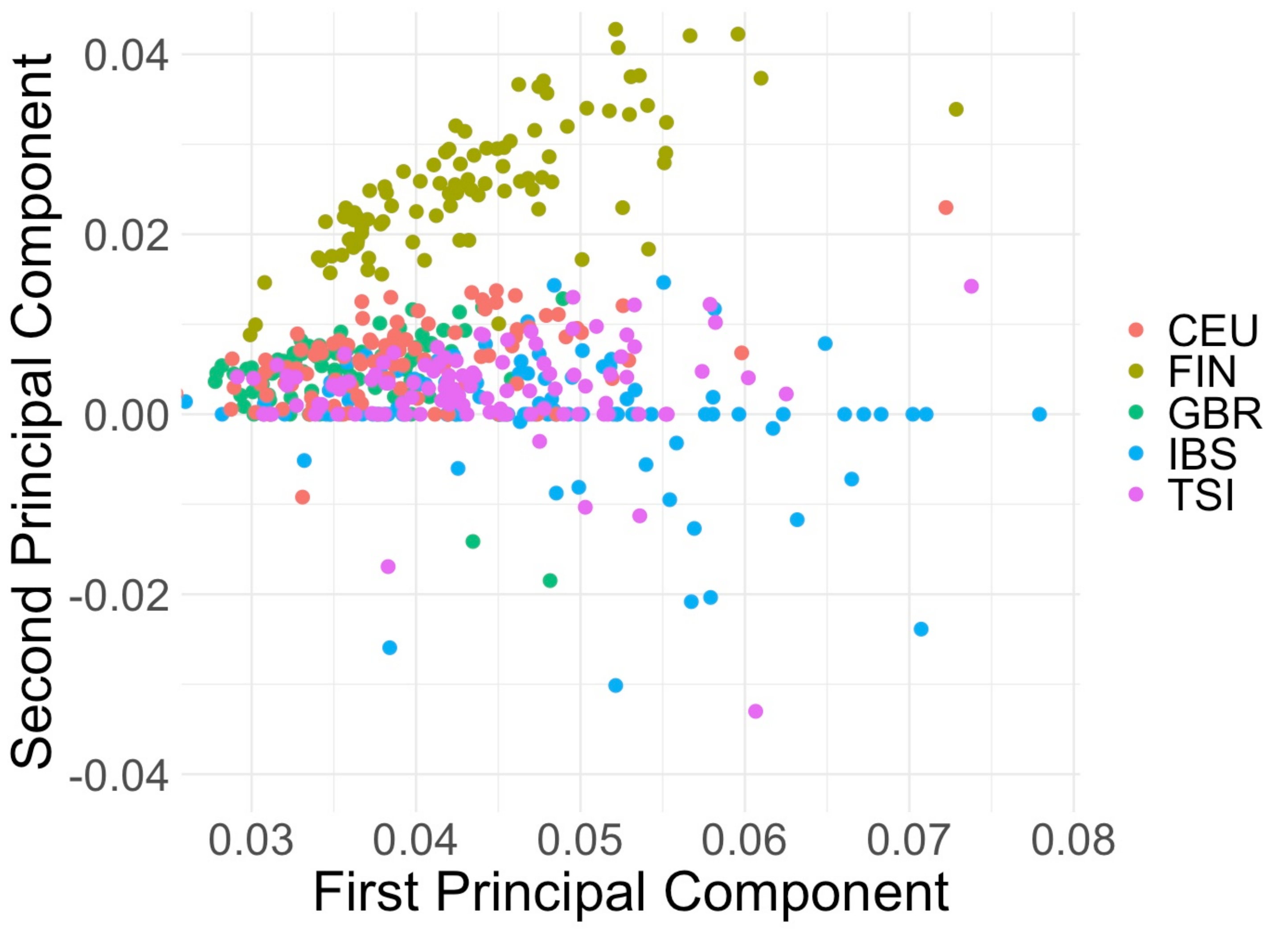}
        \caption{$\lambda=10$}
        \label{fig:unsmooth_c}
    \end{subfigure}
    \hfill
    \begin{subfigure}[b]{0.49\textwidth}
        \centering
        \includegraphics[width=\textwidth]{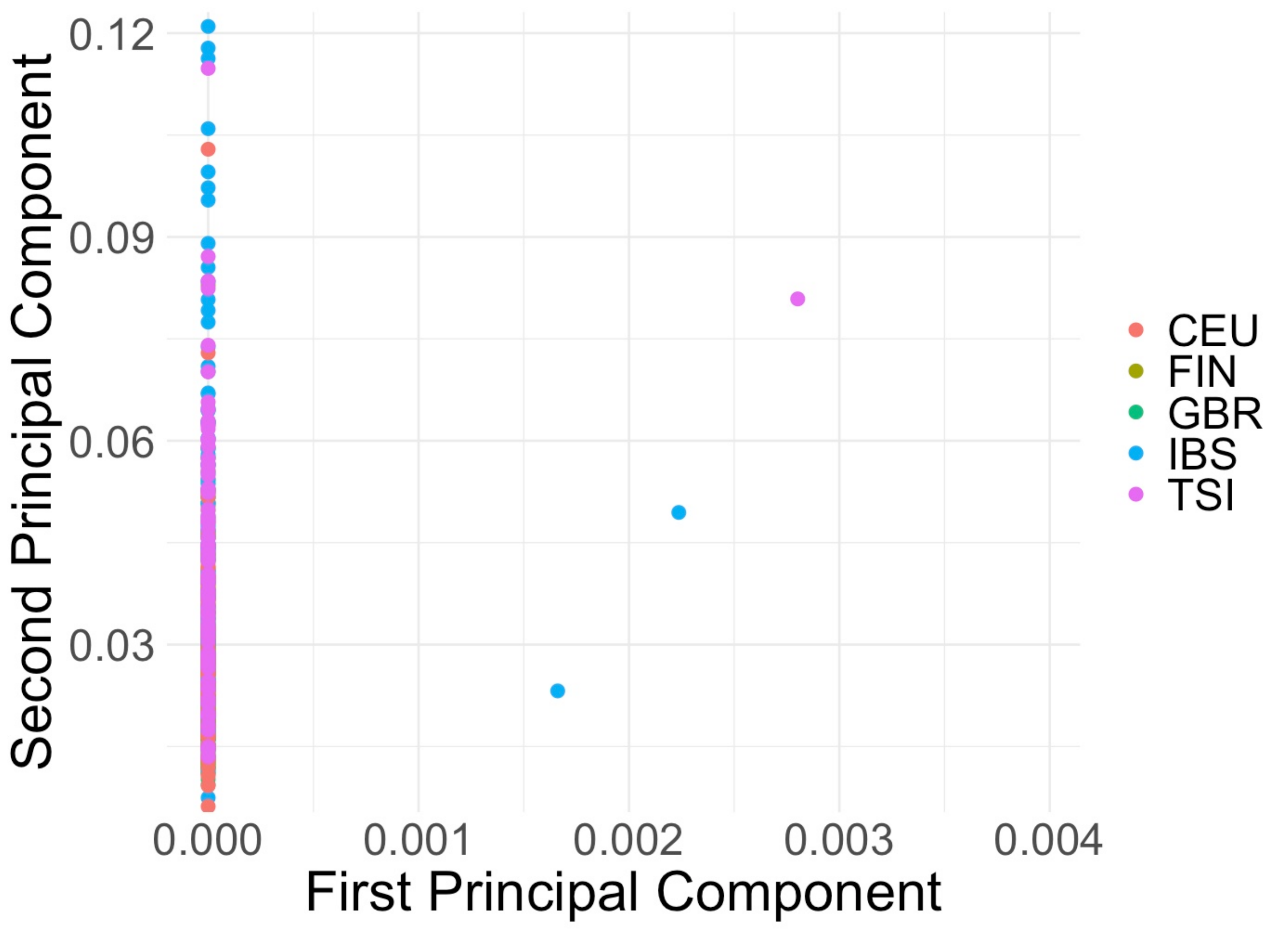}
        \caption{$\lambda=100$}
        \label{fig:unsmooth_d}
    \end{subfigure}

    \caption{Population stratification for the 1000 Genomes Project data set using unsmoothed PEP of eq.~\eqref{eq:pep}, evaluated at $\lambda \in \{0,1,10,100\}$.}
    \label{fig:unsmooth_all}
\end{figure*}

Figure~\ref{fig:unsmooth_all} displays the results of unsmoothed PEP in eq.~\eqref{eq:pep} using the first two eigenvectors, colored by subpopulation (CEU, FIN, GBR, IBS, and TSI) of the 1000 Genomes Project dataset. We evaluated unsmoothed PEP for $\lambda$ equal to 0, 1, 10, and 100. We can see a good stratification for $\lambda$ equal to 0, 1, and 10, indicated by the relative discernibility of colored population clusters (notably becoming less discernable as $\lambda$ increases). We also observe that some entries of the principal component vectors are zero, leading to points which are located on the x-axis or y-axis. The amount of entries shrunk to zero seems to increase with $\lambda$, as expected. However, when $\lambda$ increases to 100, the populations are no longer discernible.

In Appendix~\ref{sec:additional_simulations} we present two additional figures showing population stratification plots for the dataset of the 1000 Genomes Project. Those two additional figures are computed with (a) the classic eigenvectors, as well as (b) the first two principal components as in eq.~\eqref{eq:pep}, however with an $L_2$ norm instead of the $L_1$ norm.

\subsection{Results for the smoothed penalized eigenvectors}
\label{sec:smooth}
\begin{figure*}
    \begin{subfigure}[b]{0.5\textwidth}
        \centering
        \includegraphics[width=\textwidth]{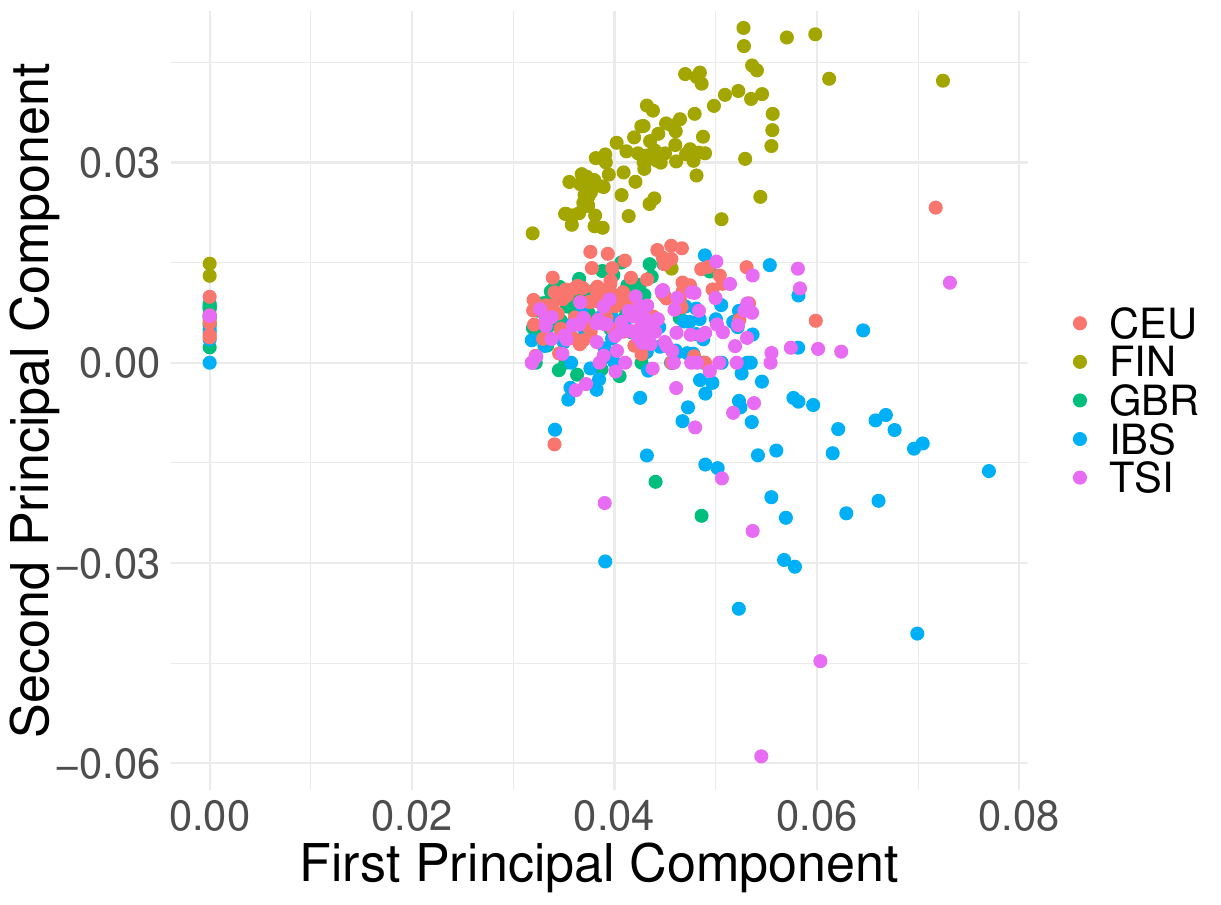}
        \caption{$\lambda=0$}
        \label{fig:smooth_0}
    \end{subfigure}
    \hfill
    \begin{subfigure}[b]{0.5\textwidth}
        \centering
        \includegraphics[width=\textwidth]{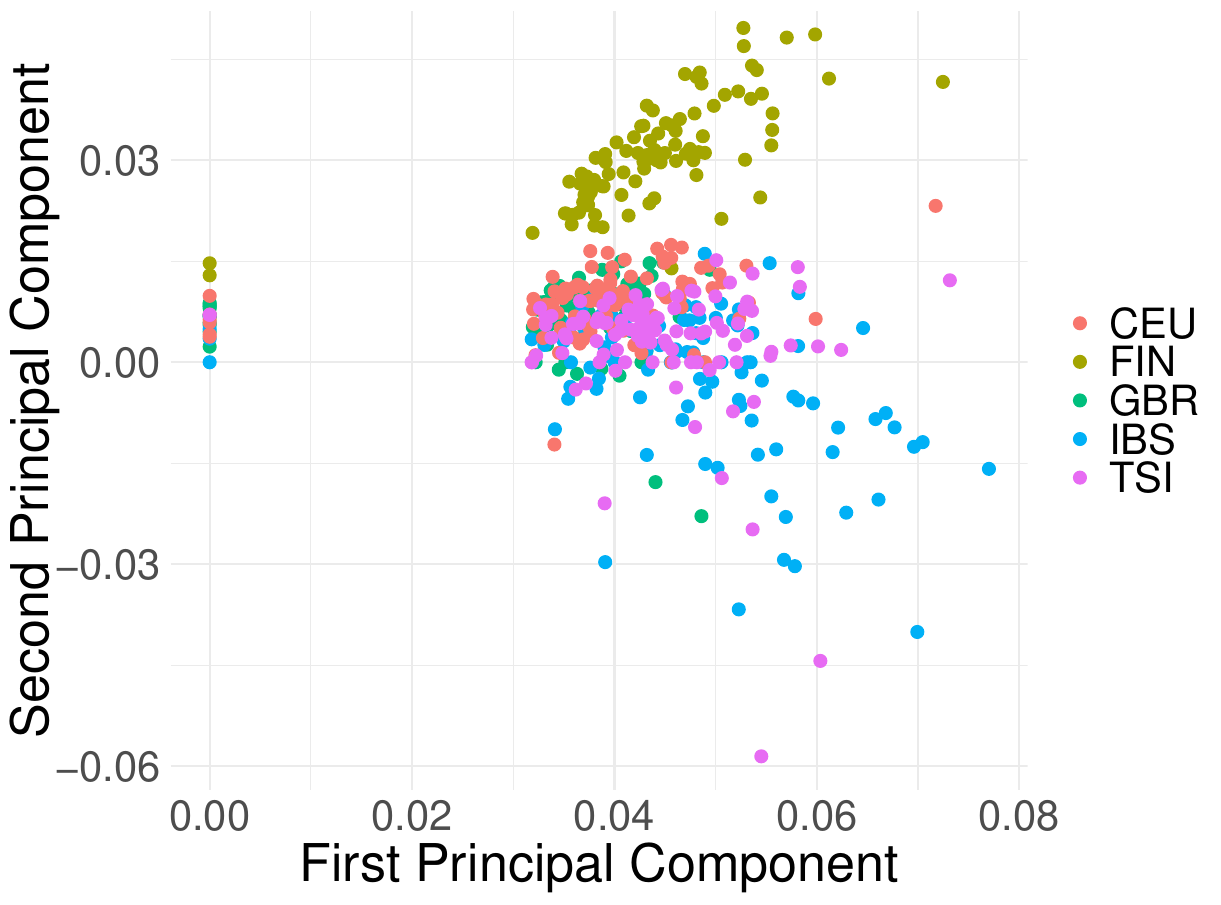}
        \caption{$\lambda=1$}
        \label{fig:smooth_b}
    \end{subfigure}
    
    \bigskip
    \begin{subfigure}[b]{0.5\textwidth}
        \centering
        \includegraphics[width=\textwidth]{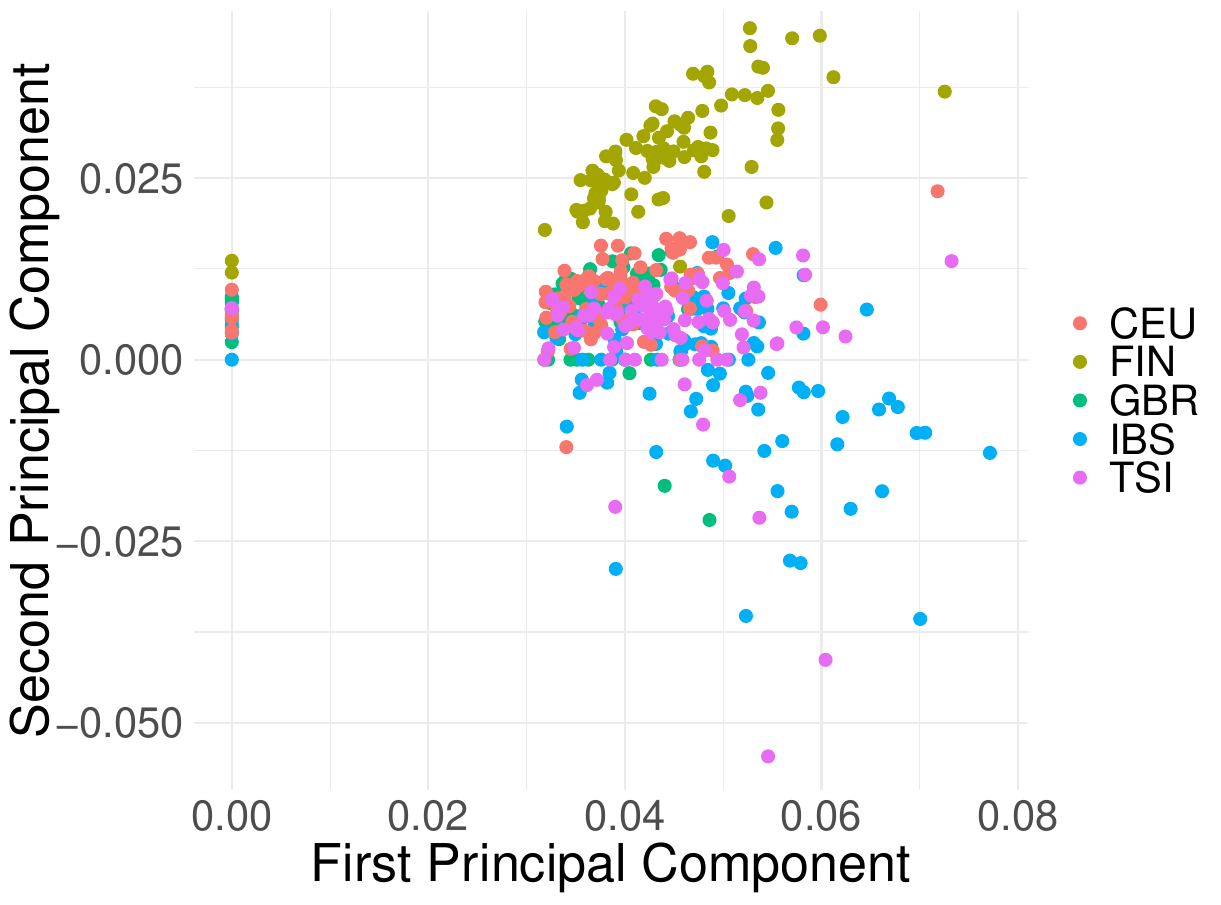}
        \caption{$\lambda=10$}
        \label{fig:smooth_c}
    \end{subfigure}
    \hfill
    \begin{subfigure}[b]{0.5\textwidth}
        \centering
        \includegraphics[width=\textwidth]{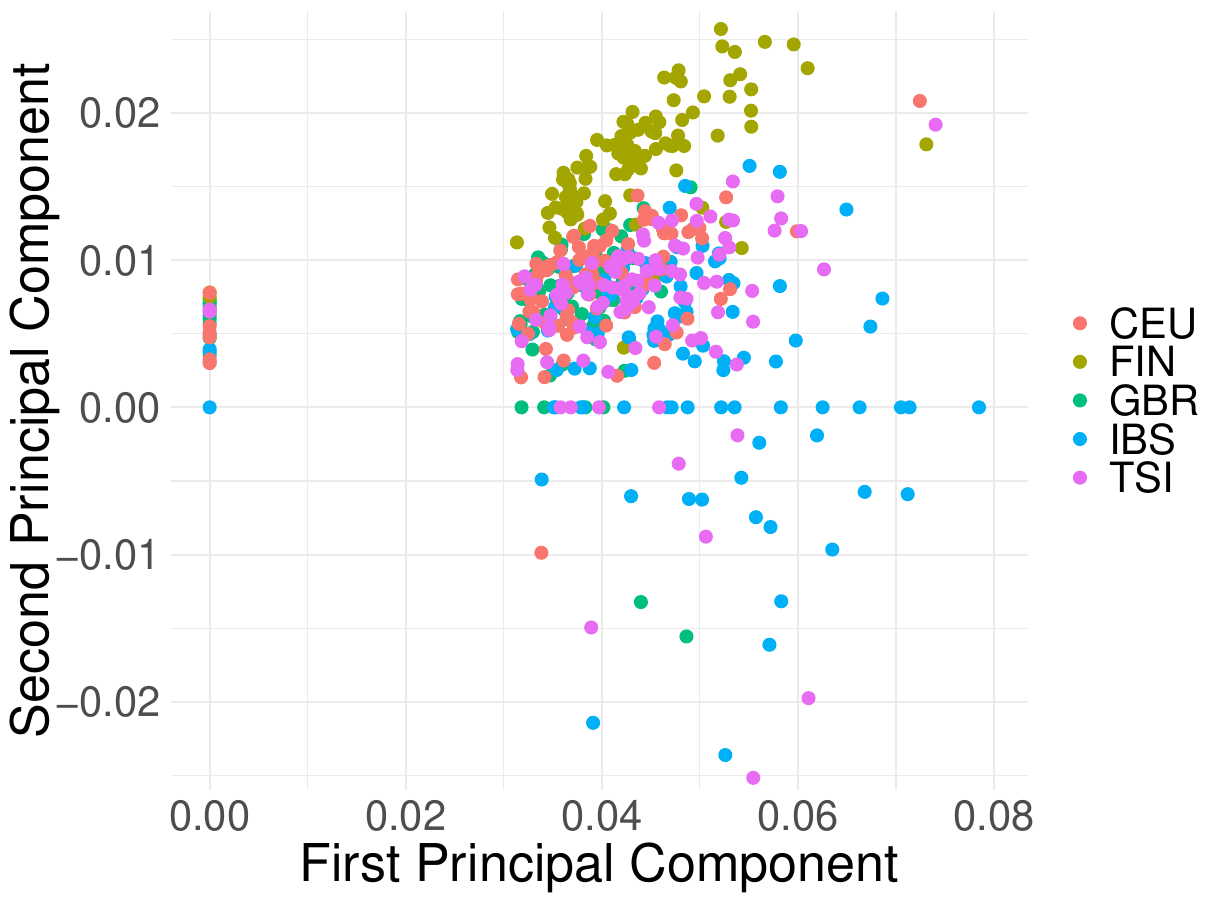}
        \caption{$\lambda=100$}
        \label{fig:smooth_d}
    \end{subfigure}
    \caption{Population stratification for the 1000 Genomes Project data set using smoothed PEP of eq.~\eqref{eq:smoothed_pep}, evaluated at $\lambda \in \{0,1,10,100\}$.}
    \label{fig:smooth_all}
\end{figure*}

Similarly to Figure~\ref{fig:unsmooth_all}, Figure~\ref{fig:smooth_all} shows the results for the smoothed version of PEP of eq.~\eqref{eq:smoothed_pep} using the first two eigenvectors, colored by subpopulation (CEU, FIN, GBR, IBS, and TSI). As before, we evaluated smoothed PEP for $\lambda$ equal to 0, 1, 10, and 100. However, in contrast to Figure~\ref{fig:unsmooth_all}, we see good stratification across all penalty values, again indicated by the relative discernibility of colored population clusters. While the clusters still become less discernable as $\lambda$ increases, for $\lambda=100$ we see a plot that is much more clearly clustered and conspicuous than the equivalent one in Figure~\ref{fig:unsmooth_all}. As before, the thresholding operation shrinks some of the entries in the first and second principal components to zero, leading to points which are located on the x-axis or y-axis. The amount of entries shrunk to zero seems to increase with $\lambda$, as expected.

\subsection{Assessing clusters}
\label{sec:clusters}
The population stratification plots of Section~\ref{sec:unsmooth} and Section~\ref{sec:smooth} exhibit slightly different clusters by population. Naturally, our aim is to achieve a good separation between the clusters. In order to quantify the quality of the clustering achieved, we compute three metrics, the within- and between-cluster sums of squares as well as the silhouette score. As in Section~\ref{sec:setting}, $n$ denotes the number of subjects, which is also the dimension of the matrix $Q \in \R^{n \times n}$ of Section~\ref{sec:methods}. We denote the number of clusters with $k$, and the set of points contained in the $i$-th cluster with $C_i$, where $\cap_{i=1}^k C_i = \emptyset$ and $\cup_{i=1}^k C_i = \{1,\ldots,n\}$. Moreover, we denote each point in the population stratification plots with $v_i \in \R^2$ and the cluster this point belongs to with $c(i) \in \{1,\ldots,k\}$, where $i \in \{1,\ldots,n\}$. The three aforementioned metrics are defined as follows:
\begin{enumerate}
    \item The within-cluster sum of squares is defined as
    \begin{align}
        \text{SS}_\text{within} = \sum_{i=1}^n \Vert v_i - \mu_{c(i)} \Vert_2^2,
        \label{eq:wss_general}
    \end{align}
    where $\mu_j = \frac{1}{|C_j|} \sum_{r \in C_j} v_r$ denotes the mean of all the points in the $j$-th cluster $C_j$, where $j \in \{1,\ldots,k\}$, and $|C_j|$ denotes the number of elements in the set $C_j$.
    \item The between-cluster sum of squares is defined as
    \begin{align}
        \text{SS}_\text{between} = \sum_{i=1}^k |C_i| \cdot \Vert \mu_i - \mu \Vert_2^2,
        \label{eq:bss_general}
    \end{align}
    where $\mu = \frac{1}{n} \sum_{i=1}^n v_i$ is the mean of all the points $v_i$, where $i \in \{1,\ldots,n\}$.
    \item The silhouette score for point $v_i$ is defined as
    \begin{align}
        \text{Silhouette}_i = \frac{b_i - a_i}{\max \{a_i,b_i\}},
        \label{eq:silhouette_general}
    \end{align}
    where $a_i = \frac{1}{|C_{c(i)}|-1} \sum_{r \in C_{c(i)}, r \neq i} \Vert v_i - v_r \Vert_2$ is the average distance of $v_i$ to all the other data points in the same cluster $C_{c(i)}$ as $v_i$, and the quantity $b_i = \min_{j \neq c(i)} \frac{1}{|C_j|} \sum_{r \in C_j} \Vert v_i - v_r \Vert_2$ is the minimum among all the mean distances of $v_i$ to the other clusters $j \neq c(i)$ that $v_i$ does not belong to. We compute the silhouette score for each point with the help of the function \textit{silhouette} of the R-package \textit{cluster} \citep{cluster}. We report the mean over all silhouette scores as a summary metric.
\end{enumerate}

By evaluating $\text{SS}_\text{within}$, $\text{SS}_\text{between}$, and $\text{Silhouette}_i$ for the projected data points, we assess the effectiveness of the penalized eigenvalue problem solutions in separating clusters in meaningful ways. These metrics provide a robust way to compare the results obtained from the unsmoothed PEP of Section~\ref{sec:unsmooth} and the smoothed PEP of Section~\ref{sec:smooth}.

Tables~\ref{tab:wss}, \ref{tab:bss}, and \ref{tab:asw} show the within-cluster sum of squares, between-cluster sum of squares, and silhouette scores for the clusters displayed in Figure~\ref{fig:unsmooth_all} and Figure~\ref{fig:smooth_all} (again for $\lambda \in \{0,1,10,100\}$). For the within-cluster sum of squares and the silhouette score (Tables~\eqref{tab:wss} and \eqref{tab:asw}), a smaller value indicates better clustering, and for the between-cluster sum of squares metric (Table~\eqref{tab:bss}), a larger value indicates a better clustering. Across all tables, for both the smoothed and unsmoothed PEP, we can see that as we increase the Lasso penalty $\lambda$, the quality of the clusters decreases across the board. We also note that as the $\lambda$ values increase, the smoothed models achieve better clustering in comparison to the unsmoothed models across all three metrics. Therefore, we posit that this smoothing method enhances the performance of these models via more discernable clustering.

\begin{table}
\caption{Within Sum of Squares clustering metric evaluating the unsmoothed and smoothed penalized eigenvector problem (PEP). Smaller within sum of squares indicates better clustering.}
\label{tab:wss}
    \begin{tabular}{lccccc}
        \midrule
        & \multicolumn{4}{c}{$\lambda$} 
        \\\cmidrule{2-5}
        Model & 0 & 1 & 10 & 100 \\
        \midrule
        Unsmoothed PEP & 1.6654 & 1.6702 & 1.7244 & 1.8839 \\
        Smoothed PEP & 1.6621 & 1.6626 & 1.6665 & 1.6910 \\
        \midrule
    \end{tabular}
\end{table}

\begin{table}
\caption{Between Sum of Squares clustering metric evaluating the unsmoothed and smoothed penalized eigenvector problem (PEP). Larger between sum of squares indicates better clustering.}
\label{tab:bss}
    \begin{tabular}{lccccc}
        \midrule
        & \multicolumn{4}{c}{$\lambda$} 
        \\\cmidrule{2-5}
        Model & 0 & 1 & 10 & 100 \\
        \midrule
        Unsmoothed PEP & 0.1278 & 0.1231 & 0.0912 & 0.0770 \\
        Smoothed PEP & 0.1334 & 0.1313 & 0.1149 & 0.0531 \\
        \midrule
    \end{tabular}
\end{table}

\begin{table}
\caption{Average silhouette score evaluating the unsmoothed and smoothed penalized eigenvector problem (PEP). Larger silhouette score indicates better clustering.}
\label{tab:asw}
    \begin{tabular}{lccccc}
        \midrule
        & \multicolumn{4}{c}{$\lambda$}
        \\\cmidrule{2-5}
        Model & 0 & 1 & 10 & 100 \\
        \midrule
        Unsmoothed PEP & -0.0459 & -0.0475 & -0.0610 & -0.2066 \\
        Smoothed PEP & -0.0175 & -0.0183 & -0.0257 & -0.0828 \\
        \midrule
    \end{tabular}
\end{table}

\subsection{Choice of the smoothing parameter}
\label{sec:mu}
We investigate the effect of the smoothing parameter $\mu$ on the obtained clustering. For this, we compute the figures of Section~\ref{sec:smooth} with different choices of $\mu$. We note that if the Lasso penalty $\lambda$ in eq.~\eqref{eq:pep} is negligible, then so will be the $L_1$ norm penalty for the objective function in eq.~\eqref{eq:pep}. Therefore, exchanging the $L_1$ norm for the smoothed counterpart will likely have a minor effect only. Therefore, we choose a relatively large Lasso penalty of $\lambda=1$ in order to see the influence of the $L_1$ penalty and, accordingly, of its smoothed approximation of Section~\ref{sec:smoothing}.

\begin{figure*}
    \centering
    \includegraphics[width=0.8\textwidth]{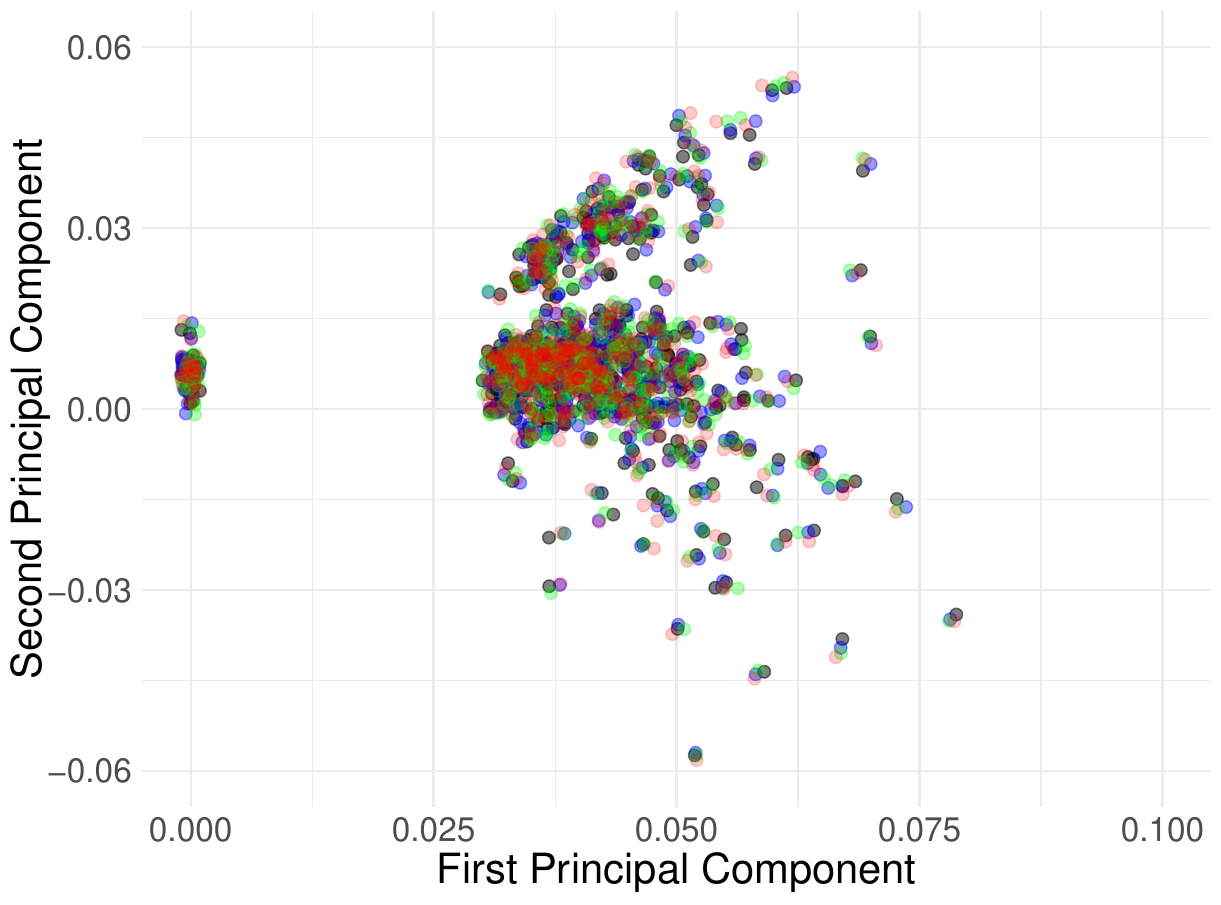}
    \caption{Population stratification for the 1000 Genomes Project data set using smoothed PEP of eq.~\eqref{eq:smoothed_pep} with Lasso penalty $\lambda=1$. Varying smoothing parameter $\mu \in \{10^{-2},10^{-1},1,10\}$ encoded with black ($\mu = 0.01$), blue ($\mu = 0.1$), green ($\mu = 1$), and red ($\mu = 10$). Log scale on both axes computed as $\log(1+x)$ and $\text{sign}(y) \log(1+|y|)$.}
    \label{fig:smooth_mus}
\end{figure*}

Figure~\ref{fig:smooth_mus} shows results for the fixed choice $\lambda=1$ and varying values of $\mu \in \{0.01, 0.1, 1, 10\}$. In contrast to Figure~\ref{fig:unsmooth_all} and Figure~\ref{fig:smooth_all}, the points in Figure~\ref{fig:smooth_mus} are not colored by population label in the 1000 Genomes Project data but unison by value of $\mu$, precisely black for $\mu = 0.01$, blue for $\mu = 0.1$, green for $\mu = 1$, and red for $\mu = 10$. We observe that as $\mu$ increases, the plots become more and more compressed in one spot, whereas smaller values of $\mu$ nicely stratify the data. We explain this finding with the fact that for a fixed $\lambda$, larger values of $\mu$ will excessively smooth out the objective function, thus rendering the maximum delocalized. In contrast, smaller values of $\mu$ cause the smoothed objective of eq.~\eqref{eq:smoothed_pep} to be pointwise close to the unsmoothed one of eq.~\eqref{eq:pep}, thus recovering the original principal components.

\subsection{Polygenic risk score}
\label{sec:prs}
We aim to compare the impact of using either the unsmoothed or the smoothed PEP eigenvectors to compute a polygenic risk score to predict SARS-CoV-2 mortality. To be precise, using the dataset described in Section~\ref{sec:setting}, we fit a linear regression model to the outcome (mortality as binary vector) as a function of age, sex, geographic region, and $10$ principal components computed on the GRM (genomic relationship matrix) similarity measure of the SARS-CoV-2 nucleotide sequences.

While keeping the regression model unchanged otherwise, we once use $10$ principal components computed with the unsmoothed PEP of eq.~\eqref{eq:pep} and smoothed PEP of eq.~\eqref{eq:smoothed_pep}. For smoothed PEP we employ $\lambda=0.1$. We train the model on a proportion of $\pi \in \{0.1,\ldots,0.9\}$ of the individuals, and evaluate the prediction against the true outcome on the withheld proportion of $1-\pi$ individuals. As the response is binary, we employ the AUC (area under the curve) metric to assess the accuracy of the prediction.

\begin{figure*}
    \centering
    \includegraphics[width=0.78\textwidth]{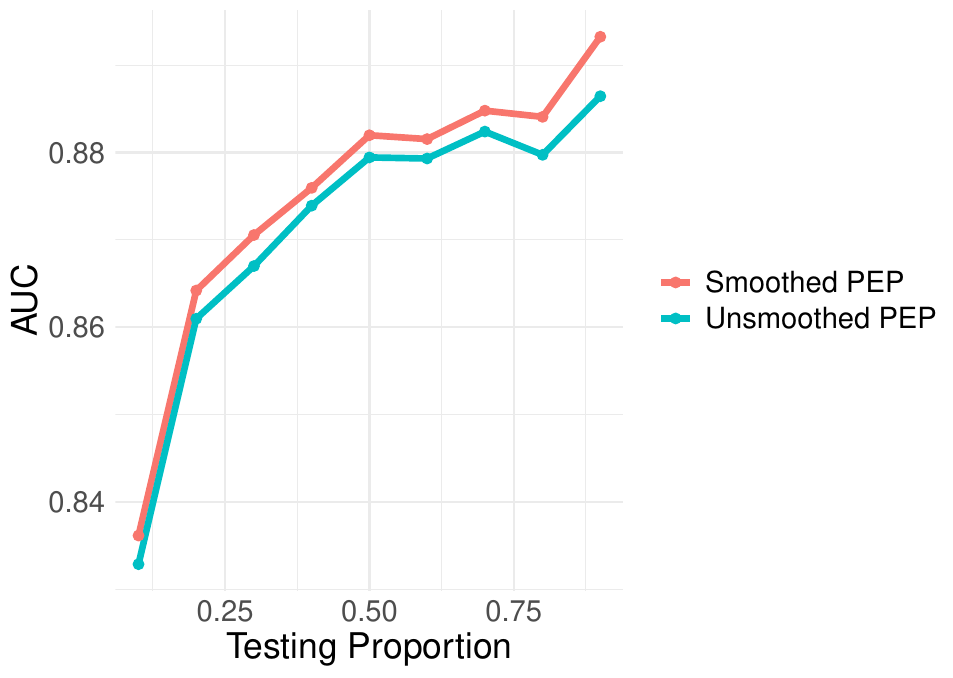}
    \caption{AUC metric for the prediction of SARS-CoV-2 mortality as a function of the proportion of the data used for training.}
    \label{fig:auc}
\end{figure*}

Figure~\ref{fig:auc} shows the results of this experiment as a function of the proportion $\pi$ used for training, both for the model using $10$ principal components from the unsmoothed (blue) and the smoothed (red) PEP. With an AUC above $0.8$, we observe that indeed, the covariates age, sex, geographic region, and the $10$ principal components of the nucleotide data are good predictors of mortality. Moreover, we observe that the prediction accuracy increases with an increasing training proportion, which is to be expected. Importantly, using the principal components from the smoothed PEP seems to give a slight edge in prediction in comparison to using the unsmoothed PEP eigenvectors.

\subsection{Clustering of the Iris machine learning benchmark}
\label{sec:iris}
As a second real data example, we aim to visually assess the clustering obtained on the Iris benchmark dataset of the UC Irvine Machine Learning Repository \citep{iris}. This simple but widely used benchmark for clustering contains data on $150$ plants, each belonging to one of three species (setosa, versicolor, virginica). The covariates are the plant's sepal length, sepal width, petal length, and petal width.

\begin{figure*}
    \centering
    \includegraphics[width=0.8\textwidth]{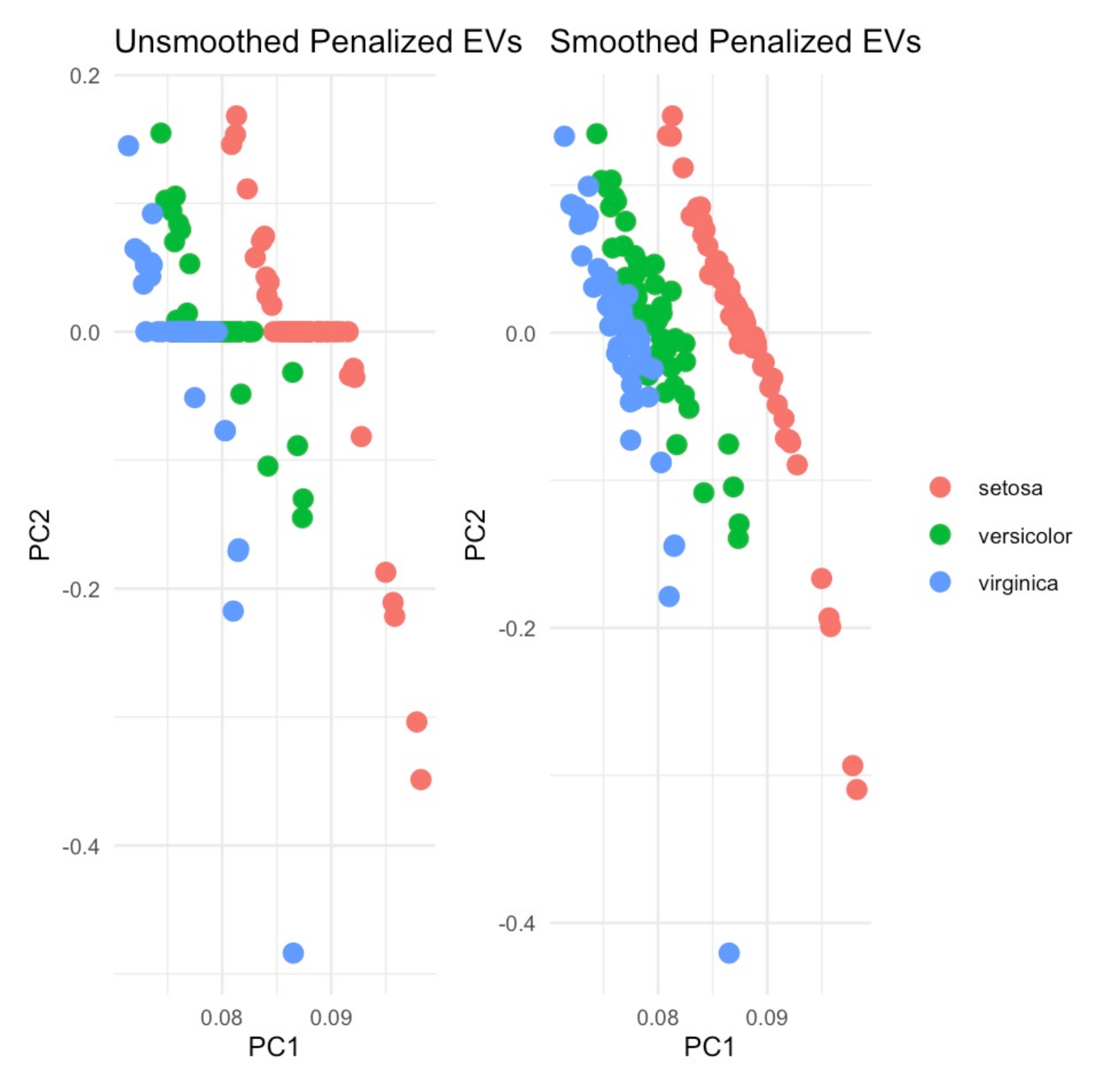}
    \caption{Iris benchmark dataset of the UC Irvine Machine Learning Repository. Clustering computed with unsmoothed (left) and smoothed (right) PEP. Points are colored by their true label (the name of the species).}
    \label{fig:iris}
\end{figure*}

As described in Section~\ref{sec:setting}, we first compute a similarity measure on the dataset to obtain a similarity matrix of dimensions $150 \times 150$, where each matrix entry is a measure of similarity between any pair of plants. As a similarity measure we use the Jaccard similarity matrix computed on $X$ \citep{Jaccard1901, Prokopenko2016, locStra}, where $X \in \R^{150 \times 4}$ is the Iris dataset (150 observations and 4 covariates). Afterwards, we compute two principal components with both unsmoothed and smoothed PEP (for $\lambda=0.075$) and plot them. The results are shown in Figure~\ref{fig:iris}, where each point is colored based on its true label. We observe that the principal components computed with smoothed PEP lead to visibly clearer clustering by species.

\subsection{Comparison with state-of-the-art sparse PCA algorithms}
\label{sec:comparison}
We compare our smoothed PEP approach of eq.~\eqref{eq:smoothed_pep} with seven other state-of-the-art algorithms on simulated data. Those approaches are the power method
for sparse principal component analysis of \cite{Journee2008}, the semidefinite programming formulation of \cite{dAspremont2007}, the penalized orthogonal iteration of \cite{Jung2019}, the penalized formulation of \cite{Gaynanova2017}, minorization-maximization algorithm for sparse PCA of \cite{Song2015}, the first-order algorithm of \cite{Zhang2010}, and the sPCA-rSVD algorithm \cite{Shen2008}.

Specifically, we implement Algorithm~6 of \cite{Journee2008} (referred to as \textit{Journee6}), Algorithm~2 of \cite{dAspremont2007} (referred to as \textit{Aspremont2}), Algorithm~2 of \cite{Jung2019} (referred to as \textit{Jung2}), Algorithm~1 of \cite{Gaynanova2017} (referred to as \textit{Gaynanova1}), Algorithm~4 of \cite{Song2015} (referred to as \textit{Song4}), Algorithm~1 of \cite{Zhang2010}  (referred to as \textit{Zhang1}), and Algorithm~1 of \cite{Shen2008} (referred to as \textit{Shen1}). Our codes also use functionality of the R-packages \textit{RSpectra} \citep{RSpectra}, \textit{alabama} \citep{alabama}, \textit{irlba} \citep{irlba}, and
\textit{epca} \citep{epca}.

In order to be able to know the true sparse eigenvector, we employ the following simulation scenario. We generate square matrices $U$ of dimensions $n \in \{10,20,50,100\}$ with a planted first eigenvector in its first column. The sparsity $\rho$ (proportion of zeros) of the first eigenvector is varied in $\rho \in \{0.1,0.2,0.3,0.4,0.5\}$. The nonzero entries of the first eigenvector are drawn independently from a uniform distribution in $[0,1]$. The remaining columns are also randomly drawn from a uniform distribution in $[0,1]$. We also generate a diagonal matrix $D$ of dimension $n$ with randomly drawn eigenvalues, sorted in decreasing order, on its diagonal. We use both $U$ and $D$ as the inputs of an eigendecomposition, and thus generate a test matrix as $X=U D U^{-1}$.

Using $X$ constructed in the aforementioned way, we apply our smoothed PEP algorithm of Section~\ref{sec:iterative} (with $\lambda=0.1$) and the aforementioned seven state-of-the-art algorithms to recover the first sparse eigenvector of $X$. As the methods of \cite{Journee2008} and \cite{Zhang2010} return a full matrix, we only consider its first column. We assess all algorithms with respect to the accuracy of the obtained eigenvectors, their support recovery, and their runtime. This is done on the same data, that is we apply all algorithms and then return four metrics for each result. To be precise, we assess the accuracy of the returned sparse eigenvector by calculating the cosine similarity measure between the returned sparse eigenvector and the true (planted) eigenvector (ranging from $0$ to $1$, where $0$ encodes maximual disagreement of the two vectors, and $1$ encodes maximal agreement). We also consider the support recovery, meaning the ability to recover the true nonzero entries. We measure this with the true positive rate, meaning the proportion of truly recovered nonzero entries (ranging from $0$ to $1$, where $0$ is worst and $1$ is best). At the same time, we also report the false positive rate, meaning the proportion of falsely proclaimed nonzero entries (ranging from $0$ to $1$, where $0$ is best and $1$ is worst). Finally, we report the runtime of all algorithms in seconds.

The accuracy results are given in Table~\ref{tab:similarity}. We observe that, as expected, the accuracy of the solution decreases with $n$. Interestingly, we do not observe a strong dependence of the accuracy on the sparsity $\rho$. All algorithms achieve a reasonable cosine similarity with the planted true eigenvector. However, the algorithms of \cite{Gaynanova2017}, \cite{Shen2008}, and our smoothed PEP perform best.

Next, we look at the support recovery, specifically the ability to recover the true nonzero entries measured with the True Positive Rate in Table~\ref{tab:tpr}. We observe that all methods, possibly apart from the one of \cite{Shen2008} as $n$ increases, accurately recover the support in this experiment. It is noteworthy that the algorithms of \cite{Jung2019}, \cite{Gaynanova2017}, \cite{Zhang2010}, and our smoothed PEP perform particularly well in this task. At the same time, we also assess the False Positive Rate, meaning the proportion of falsely identified nonzero entries in Table~\ref{tab:fpr}. Table~\ref{tab:fpr} again shows that all methods, possibly apart from the ones of \cite{Song2015} and \cite{Shen2008}, achieve a low false positive rate.

Finally, we assess the runtime (in seconds) of all algorithms in Table~\ref{tab:runtime}. We observe most methods are able to compute the leading sparse eigenvector in a fraction of a second for all matrices under consideration. Notably, the algorithms of \cite{Journee2008} and \cite{dAspremont2007} are much slower than the others.

\begin{table*}
\resizebox{\textwidth}{!}{
\begin{tabular}{cc cccccccc}
\toprule
\multirow{2}{*}{\(n\)} & \multirow{2}{*}{Sparsity $\rho$} & \multicolumn{8}{c}{Cosine Similarity} \\
\cmidrule(lr){3-10}
 & & Journee6 & Aspremont2 & Jung2 & Gaynanova1 & Song4 & Zhang1 & Shen1 & Smoothed PEP \\
\midrule
\multirow{5}{*}{10} 
 & 0.1 & 0.9037999 & 0.5275607 & 0.11229495 & 0.9914799 & 0.92853305 & 0.32581428 & 0.9759526 & 0.9914892 \\
 & 0.2 & 0.9317490 & 0.8869066 & 0.58735246 & 0.9965534 & 0.91725763 & 0.63849359 & 0.9970452 & 0.9964332 \\
 & 0.3 & 0.9089373 & 0.3214823 & 0.19747244 & 0.9973516 & 0.90620679 & 0.13018827 & 0.9894365 & 0.9971395 \\
 & 0.4 & 0.8738096 & 0.8606155 & 0.04051950 & 0.9952860 & 0.73089078 & 0.19253621 & 0.9916363 & 0.9949463 \\
 & 0.5 & 0.8927113 & 0.9034062 & 0.08860906 & 0.9945179 & 0.91792699 & 0.27524794 & 0.9971003 & 0.9942426 \\
\midrule
\multirow{5}{*}{20} 
 & 0.1 & 0.9039006 & 0.4619756 & 0.14390288 & 0.9965611 & 0.88988985 & 0.24930905 & 0.9758550 & 0.9962799 \\
 & 0.2 & 0.6297668 & 0.4087665 & 0.17633557 & 0.9965252 & 0.90907549 & 0.09943160 & 0.9712495 & 0.9963236 \\
 & 0.3 & 0.7863718 & 0.6305140 & 0.00825861 & 0.9937772 & 0.76705991 & 0.06413241 & 0.9655523 & 0.9934047 \\
 & 0.4 & 0.7391398 & 0.8164070 & 0.23050779 & 0.9954877 & 0.90933260 & 0.27164501 & 0.9815923 & 0.9952056 \\
 & 0.5 & 0.9278953 & 0.6929346 & 0.15936494 & 0.9964539 & 0.52389778 & 0.13210300 & 0.9731384 & 0.9958513 \\
\midrule
\multirow{5}{*}{50} 
 & 0.1 & 0.8838522 & 0.1805368 & 0.12775405 & 0.9921327 & 0.84785138 & 0.10742227 & 0.8802893 & 0.9920830 \\
 & 0.2 & 0.8770866 & 0.2299920 & 0.22687887 & 0.9935021 & 0.84988699 & 0.15669446 & 0.8508515 & 0.9932049 \\
 & 0.3 & 0.8785249 & 0.1048715 & 0.14177232 & 0.9943568 & 0.85364046 & 0.03533143 & 0.9335731 & 0.9936751 \\
 & 0.4 & 0.2689499 & 0.5967501 & 0.11209041 & 0.9927075 & 0.06249241 & 0.16759339 & 0.9361840 & 0.9920095 \\
 & 0.5 & 0.5923197 & 0.6376264 & 0.19823132 & 0.9946354 & 0.69934528 & 0.15969902 & 0.9290508 & 0.9936740 \\
\midrule
\multirow{5}{*}{100}
 & 0.1 & 0.8904958 & 0.2116873 & 0.10061459 & 0.9933355 & 0.88229786 & 0.17661137 & 0.7553414 & 0.9934581 \\
 & 0.2 & 0.1882329 & 0.2977159 & 0.11584467 & 0.9927417 & 0.88426056 & 0.06979850 & 0.8078601 & 0.9924776 \\
 & 0.3 & 0.3219067 & 0.1805002 & 0.06364077 & 0.9943353 & 0.86676384 & 0.05520514 & 0.8453197 & 0.9936151 \\
 & 0.4 & 0.1661451 & 0.4221019 & 0.02058195 & 0.9941975 & 0.81987146 & 0.08392663 & 0.8391396 & 0.9928401 \\
 & 0.5 & 0.8969307 & 0.2601693 & 0.00060988 & 0.9943816 & 0.90599720 & 0.24571967 & 0.8685434 & 0.9930091 \\
\bottomrule
\end{tabular}
}
\caption{Cosine similarity between the computed leading eigenvector from each sparse PCA algorithm and the true (planted) leading eigenvector as a function of the matrix dimension $n$ and the sparsity level $\rho$ (proportion of zeros).}
\label{tab:similarity}
\end{table*}

\begin{table*}
\resizebox{\textwidth}{!}{
\begin{tabular}{cc cccccccc}
\toprule
\multirow{2}{*}{\(n\)} & \multirow{2}{*}{Sparsity $\rho$} & \multicolumn{8}{c}{Support Recovery (True Positive Rate)} \\
\cmidrule(lr){3-10}
 & & Journee6 & Aspremont2 & Jung2 & Gaynanova1 & Song4 & Zhang1 & Shen1 & Smoothed PEP \\
\midrule
\multirow{5}{*}{10} 
 & 0.1 & 0.889 & 1.000 & 1.000 & 1.000 & 0.889 & 1.000 & 0.778 & 1.000 \\
 & 0.2 & 0.875 & 1.000 & 1.000 & 1.000 & 0.875 & 1.000 & 1.000 & 1.000 \\
 & 0.3 & 0.857 & 1.000 & 1.000 & 1.000 & 0.857 & 1.000 & 0.714 & 1.000 \\
 & 0.4 & 1.000 & 0.667 & 1.000 & 1.000 & 0.667 & 1.000 & 0.500 & 1.000 \\
 & 0.5 & 0.800 & 0.600 & 1.000 & 1.000 & 1.000 & 1.000 & 1.000 & 1.000 \\
\midrule
\multirow{5}{*}{20} 
 & 0.1 & 0.944 & 1.000 & 1.000 & 0.944 & 0.889 & 1.000 & 0.722 & 1.000 \\
 & 0.2 & 1.000 & 1.000 & 1.000 & 1.000 & 0.750 & 1.000 & 0.563 & 1.000 \\
 & 0.3 & 1.000 & 0.857 & 1.000 & 1.000 & 0.786 & 1.000 & 0.500 & 1.000 \\
 & 0.4 & 1.000 & 0.833 & 1.000 & 1.000 & 0.917 & 1.000 & 0.833 & 1.000 \\
 & 0.5 & 0.900 & 0.800 & 1.000 & 1.000 & 0.900 & 1.000 & 0.500 & 1.000 \\
\midrule
\multirow{5}{*}{50} 
 & 0.1 & 0.978 & 1.000 & 1.000 & 0.978 & 0.778 & 1.000 & 0.289 & 1.000 \\
 & 0.2 & 0.975 & 1.000 & 1.000 & 1.000 & 0.850 & 1.000 & 0.275 & 1.000 \\
 & 0.3 & 0.971 & 1.000 & 1.000 & 0.971 & 0.800 & 1.000 & 0.400 & 1.000 \\
 & 0.4 & 1.000 & 0.933 & 1.000 & 1.000 & 0.233 & 1.000 & 0.500 & 1.000 \\
 & 0.5 & 1.000 & 0.960 & 1.000 & 1.000 & 0.720 & 1.000 & 0.280 & 1.000 \\
\midrule
\multirow{5}{*}{100} 
 & 0.1 & 0.989 & 1.000 & 1.000 & 0.967 & 0.767 & 1.000 & 0.167 & 1.000 \\
 & 0.2 & 0.988 & 1.000 & 1.000 & 0.913 & 0.663 & 1.000 & 0.200 & 1.000 \\
 & 0.3 & 1.000 & 1.000 & 1.000 & 1.000 & 0.700 & 1.000 & 0.229 & 1.000 \\
 & 0.4 & 1.000 & 0.967 & 0.983 & 0.983 & 0.717 & 1.000 & 0.250 & 1.000 \\
 & 0.5 & 0.980 & 0.960 & 1.000 & 1.000 & 0.760 & 1.000 & 0.340 & 1.000 \\
\bottomrule
\end{tabular}
}
\caption{True Positive Rate measuring the proportion of correctly identified nonzero elements in the estimated eigenvector for each sparse PCA method as a function of the matrix dimension $n$ and the sparsity level $\rho$ (proportion of zeros).}
\label{tab:tpr}
\end{table*}

\begin{table*}
\resizebox{\textwidth}{!}{
\begin{tabular}{cc cccccccc}
\toprule
\multirow{2}{*}{\(n\)} & \multirow{2}{*}{Sparsity $\rho$} & \multicolumn{8}{c}{Support Recovery (False Positive Rate)} \\
\cmidrule(lr){3-10}
 & & Journee6 & Aspremont2 & Jung2 & Gaynanova1 & Song4 & Zhang1 & Shen1 & Smoothed PEP \\
\midrule
\multirow{5}{*}{10} 
 & 0.1 & 0.000 & 0.000 & 0.000 & 0.000 & 1.000 & 0.000 & 1.000 & 0.000 \\
 & 0.2 & 0.000 & 0.000 & 0.000 & 0.000 & 1.000 & 0.000 & 1.000 & 0.000 \\
 & 0.3 & 0.000 & 0.000 & 0.000 & 0.000 & 1.000 & 0.000 & 1.000 & 0.000 \\
 & 0.4 & 0.250 & 0.250 & 0.000 & 0.000 & 1.000 & 0.000 & 1.000 & 0.000 \\
 & 0.5 & 0.000 & 0.400 & 0.000 & 0.200 & 1.000 & 0.000 & 1.000 & 0.000 \\
\midrule
\multirow{5}{*}{20} 
 & 0.1 & 0.000 & 0.000 & 0.000 & 0.000 & 1.000 & 0.000 & 1.000 & 0.000 \\
 & 0.2 & 0.250 & 0.000 & 0.000 & 0.250 & 1.000 & 0.000 & 1.000 & 0.000 \\
 & 0.3 & 0.167 & 0.000 & 0.000 & 0.000 & 0.833 & 0.000 & 1.000 & 0.000 \\
 & 0.4 & 0.125 & 0.000 & 0.000 & 0.125 & 1.000 & 0.000 & 1.000 & 0.000 \\
 & 0.5 & 0.000 & 0.000 & 0.000 & 0.200 & 0.200 & 0.000 & 1.000 & 0.000 \\
\midrule
\multirow{5}{*}{50} 
 & 0.1 & 0.000 & 0.000 & 0.000 & 0.000 & 1.000 & 0.000 & 1.000 & 0.000 \\
 & 0.2 & 0.000 & 0.000 & 0.000 & 0.400 & 0.900 & 0.000 & 1.000 & 0.000 \\
 & 0.3 & 0.000 & 0.000 & 0.000 & 0.267 & 1.000 & 0.000 & 1.000 & 0.000 \\
 & 0.4 & 0.050 & 0.000 & 0.000 & 0.300 & 0.650 & 0.000 & 1.000 & 0.000 \\
 & 0.5 & 0.040 & 0.000 & 0.000 & 0.160 & 0.760 & 0.000 & 1.000 & 0.000 \\
\midrule
\multirow{5}{*}{100} 
 & 0.1 & 0.000 & 0.000 & 0.000 & 0.500 & 1.000 & 0.000 & 1.000 & 0.000 \\
 & 0.2 & 0.000 & 0.000 & 0.000 & 0.250 & 0.950 & 0.000 & 1.000 & 0.000 \\
 & 0.3 & 0.033 & 0.000 & 0.000 & 0.367 & 1.000 & 0.000 & 1.000 & 0.000 \\
 & 0.4 & 0.025 & 0.000 & 0.000 & 0.200 & 0.850 & 0.000 & 1.000 & 0.000 \\
 & 0.5 & 0.000 & 0.000 & 0.000 & 0.300 & 1.000 & 0.000 & 1.000 & 0.000 \\
\bottomrule
\end{tabular}
}
\caption{False Positive Rate measuring the proportion of falsely identified nonzero elements for each sparse PCA method as a function of the matrix dimension $n$ and the sparsity level $\rho$ (proportion of zeros).}
\label{tab:fpr}
\end{table*}

\begin{table*}
\resizebox{\textwidth}{!}{
\begin{tabular}{cc cccccccc}
\toprule
\multirow{2}{*}{\(n\)} & \multirow{2}{*}{Sparsity $\rho$} & \multicolumn{8}{c}{Elapsed Time (seconds)} \\
\cmidrule(lr){3-10}
 & & Journee6 & Aspremont2 & Jung2 & Gaynanova1 & Song4 & Zhang1 & Shen1 & Smoothed PEP \\
\midrule
\multirow{5}{*}{10} 
 & 0.1 & 0.028 & 0.015 & 0.003 & 0.000 & 0.002 & 0.016 & 0.000 & 0.001 \\
 & 0.2 & 0.028 & 0.047 & 0.003 & 0.000 & 0.002 & 0.015 & 0.000 & 0.001 \\
 & 0.3 & 0.027 & 0.015 & 0.003 & 0.000 & 0.053 & 0.014 & 0.001 & 0.001 \\
 & 0.4 & 0.027 & 0.233 & 0.003 & 0.001 & 0.003 & 0.015 & 0.000 & 0.002 \\
 & 0.5 & 0.027 & 0.450 & 0.003 & 0.000 & 0.002 & 0.015 & 0.000 & 0.001 \\
\midrule
\multirow{5}{*}{20} 
 & 0.1 & 0.106 & 0.052 & 0.005 & 0.000 & 0.003 & 0.027 & 0.000 & 0.003 \\
 & 0.2 & 0.108 & 0.051 & 0.005 & 0.000 & 0.003 & 0.023 & 0.000 & 0.005 \\
 & 0.3 & 0.108 & 0.349 & 0.004 & 0.000 & 0.004 & 0.023 & 0.000 & 0.005 \\
 & 0.4 & 0.109 & 0.841 & 0.004 & 0.000 & 0.004 & 0.022 & 0.000 & 0.003 \\
 & 0.5 & 0.104 & 0.780 & 0.004 & 0.000 & 0.003 & 0.023 & 0.001 & 0.002 \\
\midrule
\multirow{5}{*}{50} 
 & 0.1 & 0.740 & 0.810 & 0.018 & 0.000 & 0.007 & 0.084 & 0.001 & 0.018 \\
 & 0.2 & 0.753 & 0.720 & 0.017 & 0.000 & 0.007 & 0.093 & 0.001 & 0.015 \\
 & 0.3 & 0.736 & 0.731 & 0.018 & 0.000 & 0.007 & 0.089 & 0.001 & 0.018 \\
 & 0.4 & 0.754 & 1.901 & 0.018 & 0.001 & 0.008 & 0.096 & 0.001 & 0.016 \\
 & 0.5 & 0.738 & 4.272 & 0.018 & 0.000 & 0.007 & 0.088 & 0.001 & 0.020 \\
\midrule
\multirow{5}{*}{100}
 & 0.1 & 3.585 & 14.089 & 0.102 & 0.001 & 0.017 & 0.353 & 0.005 & 0.106 \\
 & 0.2 & 3.837 & 13.657 & 0.102 & 0.000 & 0.016 & 0.263 & 0.005 & 0.105 \\
 & 0.3 & 3.715 & 13.425 & 0.102 & 0.000 & 0.017 & 0.262 & 0.004 & 0.113 \\
 & 0.4 & 3.705 & 23.994 & 0.102 & 0.000 & 0.016 & 0.264 & 0.005 & 0.124 \\
 & 0.5 & 3.964 & 46.213 & 0.102 & 0.000 & 0.016 & 0.341 & 0.005 & 0.118 \\
\bottomrule
\end{tabular}
}
\caption{Elapsed compute times (in seconds) for each sparse PCA algorithm as a function of the matrix dimension $n$ and the sparsity level $\rho$ (proportion of zeros).}
\label{tab:runtime}
\end{table*}

\section{Discussion}
\label{sec:discussion}
This paper investigates the suitability of the Penalized Eigenvalue Problem (PEP) for visualizing population stratification in genomic data, as opposed to using ordinary eigenvectors. Our main contribution is the application of smoothing to the $L_1$ penalty of PEP \citep{Chen1995, Nesterov2005, Trendafilov2021}, thereby facilitating the efficient optimization of the PEP objective function and the computation of higher-order eigenvectors. An algorithm to iteratively apply smoothing to increase accuracy, and a thresholding approach to enforce sparsity are presented.

In an extensive experimental study, we showcase the numerical difficulties that can arise when optimizing the PEP objective function, rooted in the non-differentiability of its $L_1$ penalty. We showcase that our proposed smoothed version of PEP retains clear population stratification, and that its principal components can be utilized to correct a linear regression as demonstrated with an example in which we predicted mortality in Covid-19 patients using their age, sex, geographic region, and $10$ principal components.

Most importantly, our simulation study with seven other state-of-the-art algorithms demonstrates that for the scenarios considered in this simulation study, our smoothed PEP approach yields a high accuracy with respect to the true (planted) sparse eigenvector, a state-of-the-art support recovery (as demonstrated with a high True Positive Rate and a low False Positive Rate), and a fast runtime. It is comparable in performance to the algorithm of \cite{Gaynanova2017}, which performs equally well in this simulation study.

This project leaves scope for further research. Most importantly, a rigorous theoretical proof demonstrating the increased numerical stability of PEP is desirable and would set the proposed methodology on a rigorous mathematical foundation. Moreover, further experiments on correcting linear regressions or association analyses on genomic data are warranted. It would be interesting in this context if the modified regression problem allowed one to detect associations not detectable with ordinary principal components, for instance through smaller p-values. Finally, the application of smoothed PEP in lieu of ordinary eigenvectors has broader applicability in fields like machine learning, quantitative finance, computational biology, and computer vision via the reduction of dimensionality and the improvement of computational efficiency.

\section*{Statements and Declarations}

\subsection*{Competing Interests}
The authors have no competing interests to declare that are relevant to the content of this article.

\subsection*{Funding}
Funding for this research was provided through Cure Alzheimer's Fund, the National Institutes of Health [1R01 AI 154470-01; 2U01 HG 008685; R21 HD 095228 008976; U01 HL 089856; U01 HL 089897; P01 HL 120839; P01 HL 132825; 2U01 HG 008685; R21 HD 095228, P01 HL132825], the National Science Foundation [NSF PHY 2033046; NSF GRFP 1745302], and a NIH Center grant [P30-ES002109].

\subsection*{Acknowledgements}
The authors thankfully acknowledge support of an ERC training grant (T42 OH008416). The authors gratefully acknowledge the contributors of the 1000 Genomes Project \citep{1000genomesURL}.

The authors gratefully acknowledge the contributors, originating and submitting laboratories of the sequences from GISAID's EpiCoV\texttrademark~Database \citep{Elbe2017, Shu2017} on which this research is based. The accession numbers of all nucleotide sequences
used in this work are given in the supplementary material.

\subsection*{Data Availability Statement}
The datasets generated and/or analysed during the current study are available in the \textit{International Genome Sample Resource repository} at the address \url{https://www.internationalgenome.org/}.

The data that support the findings of this study are publicly available in the GISAID database \citep{Elbe2017, Shu2017}, see \url{https://gisaid.org/}. The supplementary material of this manuscript contains a list of all \textit{EPI\_ISL} identifiers of the nucleotide sequences which were included in the experiments.

\appendix
\section{Two additional simulation results}
\label{sec:additional_simulations}
\begin{figure}
    \centering
    \includegraphics[width=0.5\textwidth]{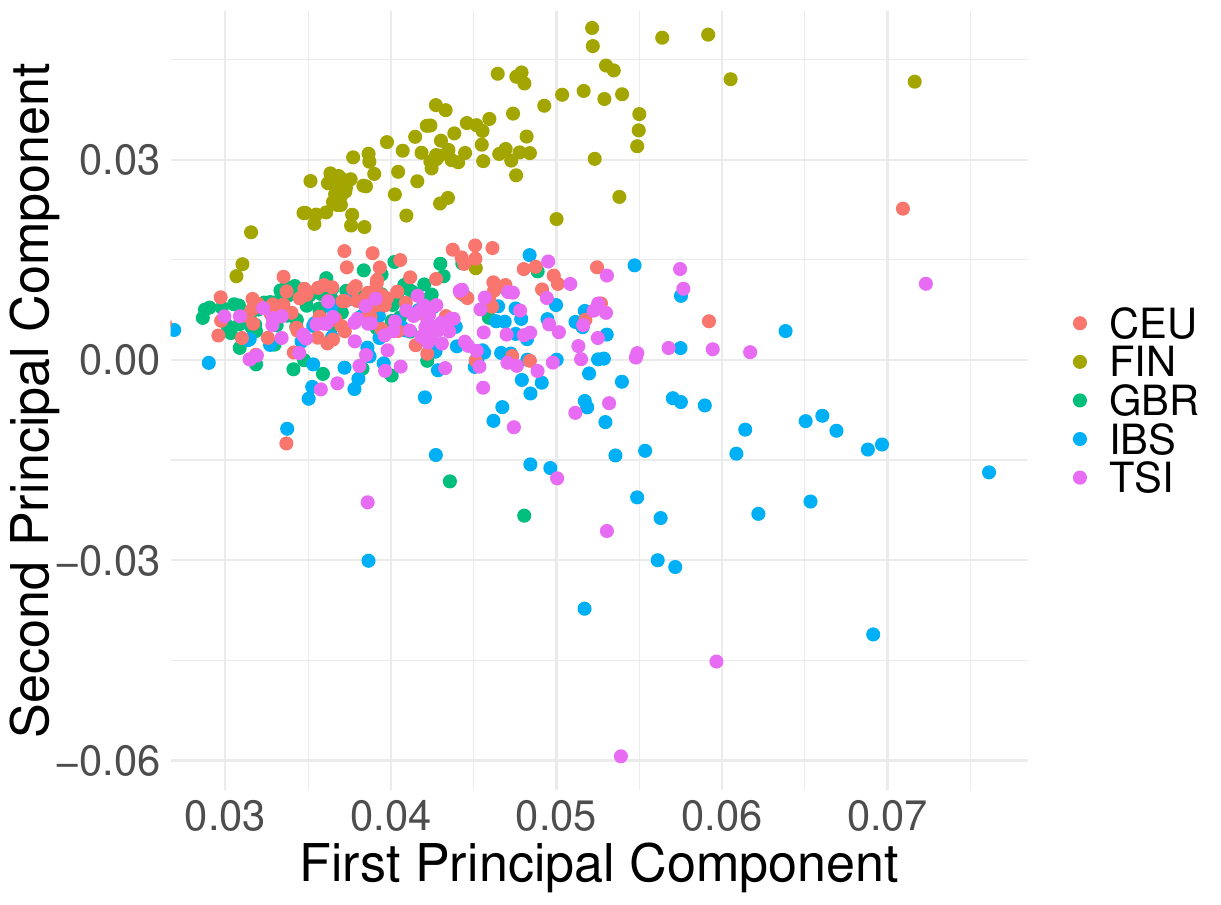}
    \caption{First two regular eigenvectors for the GRM matrix computed on the dataset of the 1000 Genomes Project. No thresholding applied.}
    \label{fig:regular}
\end{figure}

This section presents two additional simulation results. The first is a population stratification plot obtained by computing the GRM matrix on the 1000 Genomes Project dataset and by plotting its regular first two eigenvectors. The resulting plot is shown in Figure~\ref{fig:regular}. We observe that the stratification is similar to the one obtained in Figure~\ref{fig:unsmooth_all} and Figure~\ref{fig:smooth_all}, however no thresholding is applied in Figure~\ref{fig:regular}.

\begin{figure}
    \centering
    \includegraphics[width=0.5\textwidth]{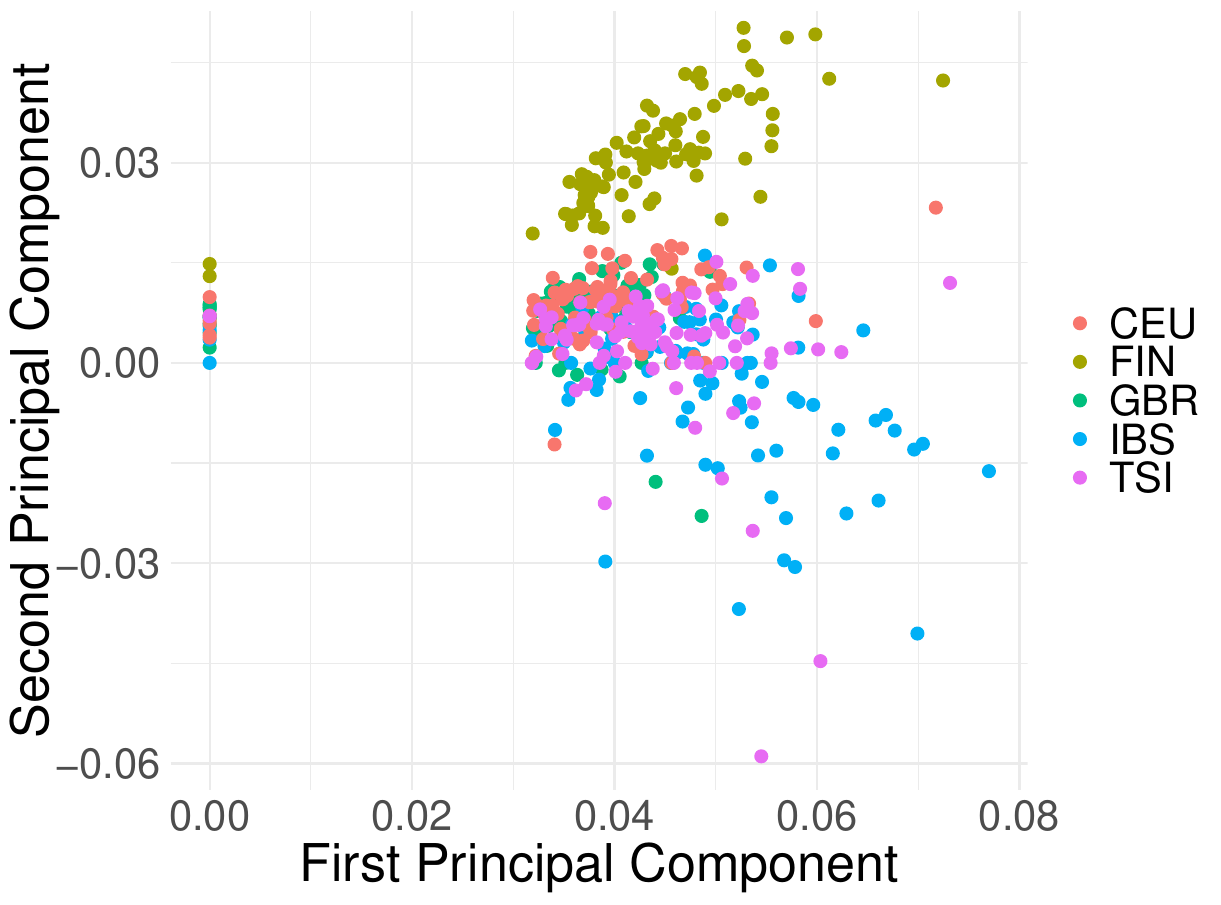}
    \caption{Principal components obtained by solving eq.~\eqref{eq:pep} with an $L_2$ norm instead of the $L_1$ norm. Thresholding applied with $5\%$ target sparsity.}
    \label{fig:L2}
\end{figure}

The second figure obtained by computing the first two principal components as in eq.~\eqref{eq:pep}, however with an $L_2$ norm instead of the $L_1$ norm. The resulting plot for $\lambda=1$ can be found in Figure~\ref{fig:L2}. It is similar to the ones seen in Figure~\ref{fig:unsmooth_all} and Figure~\ref{fig:smooth_all}.

\bibliographystyle{apalike}
\bibliography{main}
\end{document}